\documentclass[fleqn,usenatbib]{mnras}

\usepackage{newtxtext,newtxmath}
\usepackage{CJKutf8}
\usepackage{soul}
\usepackage[latin, french, german, spanish, english]{babel}
\usepackage{array}

\usepackage[T1]{fontenc}
\usepackage{orcidlink}
\usepackage{amsmath}
\usepackage{lineno}
\usepackage{natbib}
\usepackage{cleveref}

\DeclareRobustCommand{\VAN}[3]{#2}
\let\VANthebibliography\thebibliography
\def\thebibliography{\DeclareRobustCommand{\VAN}[3]{##3}\VANthebibliography}

\usepackage{graphicx}	% Including figure files
\definecolor{softgreen}{RGB}{0,100,0}	% Eye-friendly dark green
\definecolor{softred}{RGB}{178,34,34}	% Eye-friendly dark red
\definecolor{softblue}{RGB}{0,128,128}	% More visible dark teal

\newcommand{\tabincell}[2]{\begin{tabular}{@{}#1@{}}#2\end{tabular}}

\title[Stellar IMF in 100~pc]{Stellar initial mass function in the 100-pc solar neighbourhood}

\author[Y. Wang et al.]{
Yu-Ting Wang (\begin{CJK*}{UTF8}{gbsn}王雨亭\ignorespacesafterend\end{CJK*}) \orcidlink{009-0007-8574-0890}, $^{1,2}$
Chao Liu (\begin{CJK*}{UTF8}{gbsn}刘超\ignorespacesafterend\end{CJK*}) \orcidlink{0000-0002-1802-6917}, $^{1,2}$ \thanks{E-mail: liuchao@nao.cas.cn}
Jiadong Li  (\begin{CJK*}{UTF8}{gbsn}李佳东\ignorespacesafterend\end{CJK*}) \orcidlink{0000-0002-3651-5482}, $^{3}$
\\
$^{1}$National Astronomical Observatories, Chinese Academy of Sciences, Beijing 100101, People's Republic of China\\
$^{2}$School of Astronomy and Space Science, University of Chinese Academy of Sciences, Beijing 100049, People's Republic of China\\
$^{3}$Max-Planck-Institute for Astronomy, Königstuhl 17, Heidelberg D-69117, Germany\\
}

\date{Accepted XXX. Received YYY; in original form ZZZ}

\pubyear{2026}

\begin{document}
\label{firstpage}
\pagerange{\pageref{firstpage}--\pageref{lastpage}}
\maketitle
% \linenumbers

\begin{abstract}

The stellar initial mass function (IMF) is among the most fundamental distributions in astrophysics, defined as the mass spectrum of stars produced in a single star-formation event. Even in the solar neighbourhood, where measurements can be conducted via star counting, disentangling the IMF from observational effects remains challenging. In this work we introduce a new parametrisation of the stellar IMF in the 100-pc solar neighbourhood, leveraging the high-precision astrometric and photometric data from \textsl{Gaia} DR3: we model the colour-magnitude diagram of the field star population while accounting for observational uncertainties, Malmquist bias, Lutz-Kelker bias, variations in the mass-luminosity relation arising from metallicity differences, and the effects of unresolved binaries. In particular, we synthesise the binary population with a process imitating the dynamical evolution observed in star clusters to enforce that all components are drawn from the same IMF, while simultaneously recovering the observed present-day mass-ratio distribution. 
We determine an averaged stellar IMF over $0.25<m<1.0~M_{\odot}$ that aligns with canonical IMFs but achieves significantly tighter constraints: $\alpha_1=0.75^{+0.06}_{-0.04}$, $\alpha_2=2.07^{+0.04}_{-0.03}$, and a break point at $m_{\mathrm{break}}=0.40^{+0.01}_{-0.01}$ $\mathrm{M_{\odot}}$.
Our inference also yields an averaged binary fraction over $0.25<m<1.0~M_{\odot}$ of approximately 26\%, and constrains the \textsl{Gaia} DR3 angular resolution to $1.11^{+0.11}_{-0.08}$ arcsec. We also provide the $\xi$-parameter for our IMF, which is $0.5070_{-0.0096}^{+0.0068}$, to facilitate direct comparison with other IMF determinations.

\end{abstract}


%%%%%%%%%%%%%%%%%%%%%%%%%%%%%%%%%%%%%%%%%%%%%%%%%%

%%%%%%%%%%%%%%%%% BODY OF PAPER %%%%%%%%%%%%%%%%%%

\section{Introduction}
~\\
% \kp{1. 
% 1) Basic definition (Initial mass distribution for stars formed in a single star-formation event); 
% 2) how it is an abstract mathematical definition and actually cannot be extracted directly for any individual stellar population; 
% 3) so we should carefully define the IMF first and only compare the IMFs derived from similar definitions.}

The stellar initial mass function (stellar IMF) is one of the most important astrophysical distributions, defined as the mass spectrum of stars formed in a single star-formation event. The measurement of the stellar IMF is a challenging endeavour, as stellar masses cannot be measured directly. The IMF represents an abstract mathematical definition that cannot be extracted directly from any individual stellar population \citep[\textit{the IMF Unmeasurability Theorem},][]{kroupa2013}. Therefore, it is essential to clarify the specific definition of the IMF in each study based on different populations and systems, and comparisons between differently defined IMFs should be made with caution. 

% \jd{significance first, then discuss the similar definitions?}
% \kp{2. 
% The significance of studying the IMF: 1) its role as a sampling function in many simulations (benchmark, basic input); 
% 2) and its connection with multiple systems and physical process in the universe.}

Beyond single population systems, composite population systems exist in the universe, representing mixtures of multiple star-formation events; field stars in the solar neighbourhood provide one such example. The IMFs of composite stellar systems are termed composite IMFs, which should be regarded as the integration of stellar IMFs over space and time. The IMF for external galaxies also represents a composite IMF, referred to as the galaxy-wide IMF (gwIMF) by \citet{kroupa2024}. Composite IMFs differ from stellar IMFs, although they are correlated.

The IMF plays a crucial role in astronomical studies. First, it serves as a fundamental input, specifically a mass sampling function, for simulations in astronomy \citep[e.g., simulations of cluster dynamical evolution, galactic population synthesis, etc.][]{wang2020a, vazdekis2015}.
Moreover, the IMF provides a valuable tool for understanding physical processes of the baryon cycle in the universe. Since direct stellar mass measurements are difficult and inefficient (requiring binary orbital motion or asteroseismology), the IMF offers a more tractable means of extracting knowledge about star formation and evolution from a statistical perspective.

% \kp{3. 
% 1) Key issues in the field of IMF; and what we care about is whether the IMF is varying with star-formation conditions; 
% 2) failed to find the evidence of varying IMF in the star-resolved regions in early studies (citations); 
% 3) with the improvement of data quality, plenty of studies reported the variation of the IMF (citations), mainly based on observations of external galaxies. 
% 4) Disadvantages: represent the variation by indirect values (e.g. and citations); hard to constrain the low-mass end.}

Two key issues dominate the field of IMF research. The first concerns the nature of the IMF: specifically, why the IMF takes the form of a power law combined with several characteristic masses. The power-law form suggests a generally scale-free process during star formation, whereas the emergence of characteristic masses (turning points/break points/peaks) within the power law may indicate additional physics that provides a termination scale for the star formation cascade \citep{Guszejnov2016,gjergo2026}. 

The second is the variation of the IMF \citep{kroupa2013}. Theories of star formation predict that the IMF varies with the stellar environment \citep{bonnell2006, clark2011}. However, in early studies, star counts in star-resolving regions did not observe variations in the IMF over $3\sigma$ level \citep[but see \citealt{kroupa2026} for an up-to-date review]{natebastian2010}. As the quality of observational data improved, many studies reported evidence of IMF variations across different galactic environments recently. Since these studies focused on extragalactic systems, what they discovered are actually the variation of the gwIMFs. According to the IGIMF theory \citep{Weidner2005}, the high-mass end of the gwIMF is varying with the average star formation rate (SFR) of galaxies even when setting the stellar IMF universal. This means that, in composite populations, the stellar IMF and star formation history (SFH) are highly degenerated at high-mass end ($m\gtrsim 1.0 M_{\odot}$). Therefore, if we want to study the variation of stellar IMF in composite stellar systems, it is simpler to do it in the low-mass end ($m\lesssim 1.0 M_{\odot}$) where stars hardly evolve during their lifetime, and where, assuming a universal lower mass limit for the IMF regardless of age and metallicity, the integrated gwIMF of the field population approximates the underlying stellar IMF.

% \kp{4. We are still looking forward to finding the evidence of the variation of the IMF at the low-mass end in the MW, where stars can be resolved.}

As a prerequisite, it is essential to establish a method capable of accurately measuring the stellar IMF of field stars in the Milky Way (MW). Stellar samples from the solar neighbourhood constitute a subset of the MW's field stars and provide an excellent sample for investigating the IMF of brown dwarfs and low-mass stars. In the solar neighbourhood, we can obtain the highest-quality photometric, trigonometric parallax, and spectroscopic measurements. Precise measurement and detailed analysis of the stellar IMF in the solar neighbourhood can serve as a bridge to other regions, including the broader Milky Way, star-forming regions, the Galactic centre, star clusters, and external galaxies.

In 1955, Salpeter published a seminal work on the IMF \citep{salpeter1955}, in which he described the IMF using a single power-law distribution, expressed as 
\begin{equation}
\frac{\mathrm{d}N}{\mathrm{d}m} = \xi(m) = km^{-\alpha},
\label{eq:1}
\end{equation}
where \( \mathrm{d}N \) represents the number of stars within the mass interval \( \mathrm{d}m \). He derived the mass distribution for main-sequence stars ranging from 0.4 to 10 \( M_{\mathrm{\odot}} \) in the solar vicinity by combining the largest available observational samples of the nearby stellar luminosity function 
\begin{equation}
\phi(M_{\mathrm{V}}) = \frac{\mathrm{d}N}{\mathrm{d}M_{\mathrm{V}}}
\label{eq:2}
\end{equation}
available at the time. This resulted in determining the slope of the mass distribution as \( \alpha = 2.35 \), a relation he termed the ``original mass function'' for stars. In \citet{salpeter1955}, the single power-law IMF is referred to as the Salpeter IMF, and any IMF with a single power-law index \( \alpha \) close to $2.35$ (or \( \Gamma=1-\alpha = -1.35\)) is said to possess the ``Salpeter slope''.

Subsequently, pencil-beam survey-technique was developed, focusing on small field of view ($<5~\mathrm{deg}^2$), typically for extragalactic scientific purposes. Research on the IMF began to focus on two types of samples: nearby ($5.2~\mathrm{pc}$) trigonometric parallax samples \citep{jahreiss1997} and deeper pencil-beam stellar count data \citep{reidgilmore1982,gilmorereid1983,gouldDiskDwarfLuminosity1996, zheng2001}. \citet{miller1979} calculated the IMF for stars with masses less than 1.0 $M_{\odot}$ using nearby stellar samples, while \citet{scalo1986} primarily relied on more recent pencil-beam observational samples.

In the 1990s, improved understanding of the non-linearity in the stellar mass-luminosity relation (MLR) and detailed analysis of the effects of unresolved binary systems \citep{kroupa1993} led to revisions of the IMF for low-mass stars. For the first time, the shape of the luminosity function was explained from the perspective of stellar physics, and all biases in stellar count data based on trigonometric and photometric parallaxes were rigorously modelled. This addressed the discrepancies between results obtained by \citet{miller1979} and \citet{scalo1986}, specifically the differences between luminosity functions and mass functions derived from the 5.2 pc trigonometric parallax sample and the photometric parallax samples from pencil-beam surveys directed towards the Galactic poles. \citet{kroupa2001} conducted a comprehensive compilation and analysis of IMF determinations from young star clusters, OB associations, and star-forming regions. These studies provided consistent IMF results, which are referred to as the ``Kroupa IMF'', characterized by the power-law indices of a piece-wise power-law distribution:

\begin{equation}
\xi(m) =
\begin{cases}
m^{-1.3\pm 0.5}, & 0.08 < m/\mathrm{M_{\odot}} \leq 0.5 \\
m^{-2.3\pm 0.3}, & 0.5 < m/\mathrm{M_{\odot}} \leq 1.0 \\
m^{-2.3\pm 0.7}, & m/\mathrm{M_{\odot}} > 1.0
\end{cases}
\label{eq:3}
\end{equation}

\citet{reidPalomarMSUNearby2002} utilized updated trigonometric parallax data from the Hipparcos mission to confirm the results of \citet{kroupa1993,kroupa2001}. \citet{chabrierGalacticDiskMass2003} refitted the mass function for the mass range of 0.07 to 1.0 \(\mathrm{M_{\odot}}\) using a log-normal function for the lower mass end, while retaining the ``Salpeter IMF'' form for the higher mass end. This bimodal mass function, characterized by a log-normal distribution with a mean (\(\mu\)) of 0.08 and a standard deviation (\(\sigma\)) of 0.68 at the lower mass end and by the Salpeter form at the higher mass end, was subsequently termed the ``Chabrier IMF''. Despite the difference in mathematical forms, the Chabrier IMF is statistically indistinguishable from the Kroupa IMF \citep{kroupa1993}.

In the 2000s, large-scale digital sky surveys such as the Two Micron All Sky Survey \citep[2MASS;][]{skrutskieTwoMicronAll2006} and the Sloan Digital Sky Survey \citep[SDSS;][]{york2000} emerged, providing more uniform and deeper photometric imaging across extensive areas of the sky. \citet{covey2008} compiled photometric catalogues from SDSS, 2MASS, and other surveys to study the present-day mass function (PDMF) of field stars, comprising approximately 30,000 low-mass stars. \citet{bochanski2010} employed a photometric sample from the Sloan Digital Sky Survey Data Release 6, covering 8400 deg\(^2\) of the sky and comprising 15,000,000 stars, to derive photometric parallaxes and determine the luminosity function (LF) and mass function (MF) for low-mass stars in the range \(0.1 \lesssim m/\mathrm{M_{\odot}} \lesssim 0.8\), finding good consistency with previous work. However, the computation of photometric parallaxes involves the use of colour-luminosity relations derived from limited spectroscopic observations, thereby introducing multiple uncertainties.

% 比起已有的用Gaia测IMF的工作，优势？更可靠、物理的双星建模
The \textsl{Gaia} astrometric mission \citep{gaiacollaboration2018} has revolutionized the trigonometric parallax catalogue of the Milky Way. Using data from \textsl{Gaia} DR2, Sollima selected over 120,000 stars to study the IMF of the Galactic disc \citep{sollimaStellarInitialMass2019}. His model accounted for unresolved binary systems, metallicity distribution, star formation history, and the variation in stellar number density across the Galactic disc, along with all observational effects. His results are consistent with the canonical IMF within the margins of error. Subsequently, \citet{hallakoun2021} and \citet{li2023} also used \textsl{Gaia} data to study the variation of the IMF in the solar neighbourhood. However, few meticulous field-star population synthesis models have been established with detailed binary modelling in the solar neighbourhood.

% \kp{6. Our goal: consider the effect of unresolved binaries (Gaia resolution, binary population evolution), and variation of mass-luminosity relation, obtain the mass function before the dynamical evolution.}

In this study, we use the most up-to-date, largest, and most complete sample with photometric and astrometric observations in the solar neighbourhood to remeasure the stellar IMF for field stars in this region. In our model, the effects of Malmquist bias, Lutz-Kelker bias, variations in the mass-luminosity relation (MLR) for stars of different metallicities, and the effects of multiple systems are comprehensively addressed.  In \Cref{sec:data} we describe the observational data, the selection cuts applied, and how we assess the quality and completeness of the data. In \Cref{sec:method} we present our analysis methods, including clarification of basic definitions, the procedure for synthesizing the field star population, and how we use Bayesian inference to obtain posterior distributions for the parameters of interest for solar neighbourhood field stars. In \Cref{sec:res} we describe the primary results of our work. Furthermore, we investigate the reliability of the method through several tests and discuss caveats in \Cref{sec:dis}, before presenting our conclusions in \Cref{sec:conclusion}.

\section{Data} \label{sec:data}

\subsection{\textsl{Gaia} Early Data Release 3: The \textsl{Gaia} Catalogue of Nearby Stars (GCNS)} \label{sec:GCNS}

% \kp{check for the quality of the photometry and astrometry, corrections}

In this work, we utilize the catalogue from the \textsl{Gaia} Early Data Release 3 (EDR3), which contains all objects with reliable astrometry and non-zero probabilities of being within 100~pc of the Sun; this will be referred to as the \textsl{Gaia} Catalogue of Nearby Stars \citep[GCNS; ][]{collaborationGaiaEarlyData2021b} hereafter.
The catalogue includes over 330,000 objects in the 100-pc solar neighbourhood. We assess the photometry, astrometry, and completeness of sources in the GCNS catalogue before proceeding to detailed analysis of the initial mass function (IMF) embedded in the data. 

For the $G$-, $G_{\rm{BP}}$- and $G_{\rm{RP}}$-band apparent magnitude provided in the catalog, three corrections/problems are taken into account \citep{rielloGaiaEarlyData2021}: 

First, systematics arise from the use of a default colour in the Image Parameter Determination (IPD) process when calculating the $G$-band fluxes in \textsl{Gaia} EDR3.
This effect has been corrected in the \textsl{Gaia} DR3 catalogue; therefore, we substitute the $G$-band magnitude with the DR3 value. 

Second, the mean $G_{\rm{BP}}$ flux for faint red sources is overestimated because epochs with a calibrated flux lower than $1~e^{-}s^{-1}$ are excluded when computing the weighted mean flux for a source. This exclusion is necessary because a source with a flux below this limit cannot be observed as following a normal distribution.
From the tests conducted by \citet{rielloGaiaEarlyData2021}, stars with $G_{\rm{BP}}<20.3~$mag are minimally affected by this issue. Thus, we apply a cut of $G_{\rm{BP}}<20.3~$mag.

Finally, we consider the crowding effect. $G_{\rm{BP}}$ and $G_{\rm{RP}}$ fluxes represent the integrated mean fluxes obtained from CCD windows of $3.5\arcsec \times 2.1\arcsec$, whereas the $G$ flux is determined from Point Spread Function (PSF) or Line Spread Function (LSF) fitting to an object window from the astrometric field CCDs \citep{rielloGaiaEarlyData2021}.
In crowded fields, sources measured in the $G_{\rm{BP}}$ and $G_{\rm{RP}}$ bands tend to appear brighter and bluer. Stars with problematic BP/RP photometry can be identified using the parameter $\beta$ defined in \Cref{eq:4}, using the columns from the \textsl{Gaia} EDR3 archive. We remove these semi-resolved sources ($\beta<0.5$), which constitute only approximately 4 per cent of the entire data set and thus cause a modest effect on completeness.

\begin{equation}
\beta=\frac{{\textmd{phot\_bp\_n\_blended\_transits} + \textmd{phot\_rp\_n\_blended\_transits}}}{{\textmd{phot\_bp\_n\_obs} + \textmd{phot\_rp\_n\_obs}}}
\label{eq:4}
\end{equation}

Regarding the astrometric information in the catalogue, we consider zero-point correction and include the NSS parallax solution.

We intend to apply parallax zero-point corrections using the official package \texttt{gaiadr3-zeropoint}, which can be applied to \textsl{Gaia} EDR3 and DR3 data \citep{lindegrenGaiaEarlyData2021a}. However, we find that the absolute change in parallax from correcting our sample is less than 1 per cent. Given the small impact of the correction, and the fact that some of our sources lack parameters required by the zero-point correction package while completeness is more critical for the star-counting approach to determining the IMF, we neglect the zero-point correction. The zero-point correction should not be neglected when the sample covers a wider distance range.

Furthermore, astrometric solutions can produce unreliable measurements for very nearby multiple systems, as the internal orbital motion of such systems can make it challenging to isolate their parallax motion accurately. Therefore, we use the non-single star (NSS) parallax solution to replace sources that suffer from this issue \citep{gaiacollaboration2023}.

\subsection{Completeness test} \label{sec:completeness}
% \jd{a new subsection? e.g., 2.2 Completeness}
% \kp{calculate the effective volume, completeness}

The completeness of a volume-limited dataset, such as the GCNS, is crucial for the derivation of the IMF. Selection effects primarily impact the absolute magnitude $M_{\rm{G}}$ and position (e.g., Galactic latitude and Galactic longitude), while the colour $G_{\rm{BP}}-G_{\rm{RP}}$ remains relatively unaffected \citep{cantat-gaudin2023}. Since the stellar distribution in the GCNS catalogue is relatively homogeneous, variations in selection effects with position can be ignored.

\begin{figure*}
    \centering
    \includegraphics[width=0.8\linewidth]{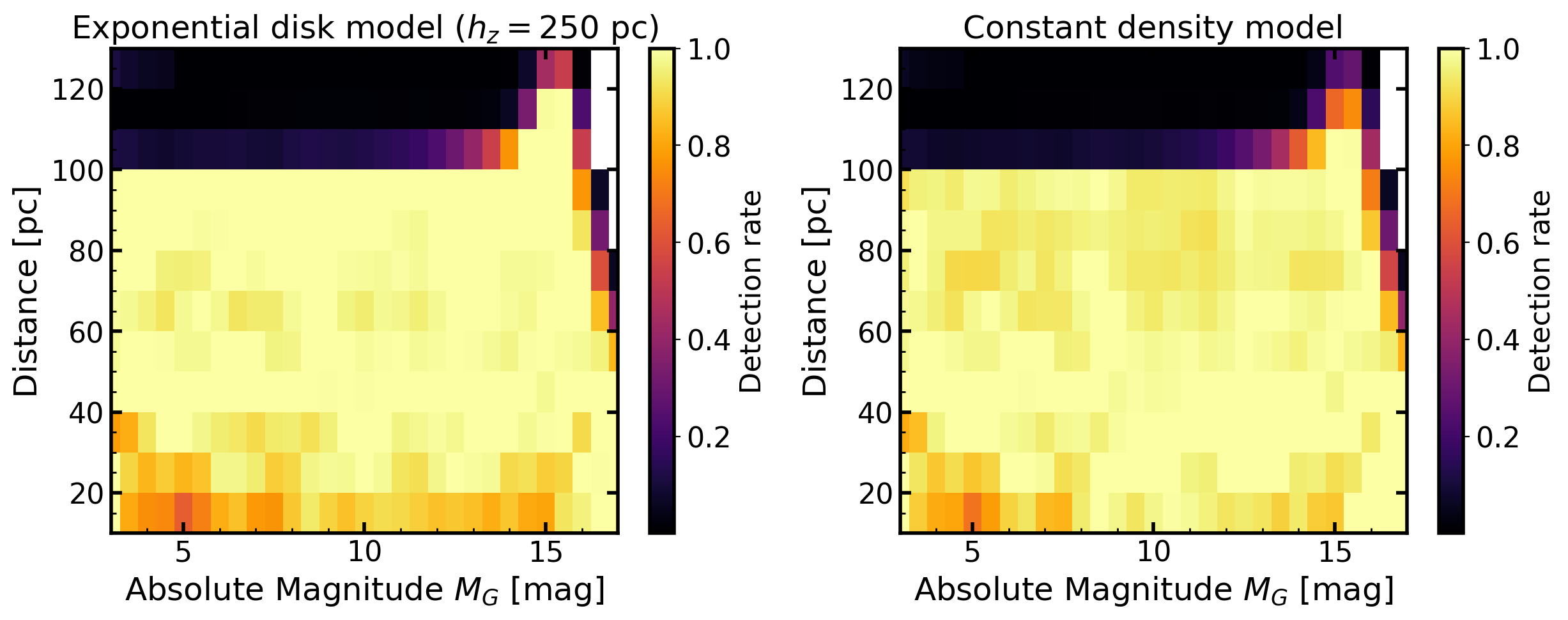}
    \caption{Detection rate (selection function) for stars within different $M_{\mathrm{G}}$ and distance ranges, assuming different density profile models. Left: exponential disk density model. Right: constant density model. See \Cref{sec:app_vmax} for details.}
    \label{fig:detec_rate}
\end{figure*}

We analyze the detection rate (completeness) of the GCNS sample following the method of \citet{rybizki2018} and \citet{mateu2020}, comparing the observed stellar counts to control samples generated with different assumed density profiles (constant and exponential disk models). Details of this analysis are provided in \Cref{sec:app_vmax}. \Cref{fig:detec_rate} shows the resulting detection rates as a function of $M_{\mathrm{G}}$ and distance. For the magnitude range $4.1<M_{\mathrm{G}}<12.1$~mag that we focus on, the sample exhibits high and uniform detection rates across most of the 100-pc volume. Only a small region corresponding to nearby bright stars (lower-left in \Cref{fig:detec_rate}) shows slightly lower detection rates, likely due to flux saturation in \textit{Gaia}'s photometry and the small number statistics of Poisson distribution leading us to a mode smaller than the truth. 
This affected region comprises only approximately 0.6\% of our main sample. Crucially, we find no systematic trend in detection rate along the $M_{\mathrm{G}}$ axis for neither density model, indicating that our sample is effectively complete for IMF determination in the range $4.1<M_{\mathrm{G}}<12.1$~mag within 100~pc. Therefore, no additional selection function corrections are applied when deriving the IMF.

\subsection{Sample selection} \label{sec:MS sample}
% \kp{how we selected out the MS stars}

\begin{figure}
    \centering
    \includegraphics[width=0.85\linewidth]{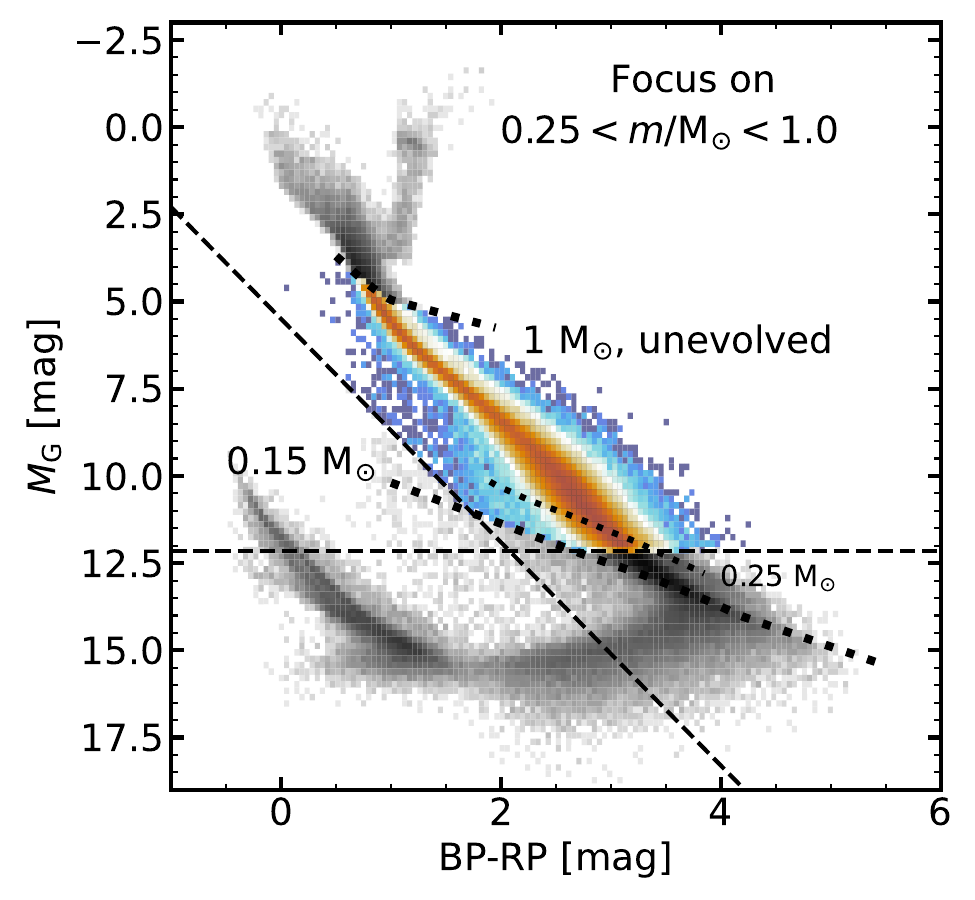}
    \caption{colour-magnitude diagram for stars in the 100-pc solar neighbourhood. The scatter with gray colour map illustrates the density distribution for all GCNS sample, overlapped with the object density distribution resulting from our sample selection procedure in \Cref{sec:MS sample}. Black dashed lines denote the selection standards for Main-sequence stars in \citet{penoyre2022}, while black dotted lines denote the iso-mass lines for $0.15~\mathrm{M_{\odot}}$ and unevolved $1.0~\mathrm{M_{\odot}}$ stars from PARSEC model.}
    \label{fig:GCNS_cmd}
\end{figure}

Next, we explain how we select stars from the GCNS catalogue for the calculation of our IMF. In \Cref{fig:GCNS_cmd}, we show the colour-magnitude diagram (CMD) for all stars in the GCNS as gray dots. We compute the absolute magnitude using the equation

\begin{equation}
M_{\mathrm{G}}=G+5\mathrm{log_{10}}(\varpi)-10,
\label{eq:5}
\end{equation}
where $M_{\mathrm{G}}$ stands for the absolute magnitude of \textsl{Gaia} G-band, $G$ the G-band apparent magnitude, and $\varpi$ the parallax measured by \textsl{Gaia}.

In this study, we focus on the initial mass function (IMF) of low-mass main-sequence stars ($0.25 < m/\mathrm{M_{\odot}} < 1.0$), which experience minimal evolutionary effects, simplifying our modelling approach. Interstellar extinction is negligible within the 100-pc solar neighbourhood \citep{edenhofer2024}, allowing us to rely on \textit{Gaia} DR3 photometry without correction. To select main-sequence stars, we adopt the boundary from \citet{penoyre2022}, defined as $M_{\mathrm{G}} - 3.2(G_{\mathrm{BP}} - G_{\mathrm{RP}}) < 3.8$. For the low-mass end, we use PARSEC iso-mass lines at $0.15~\mathrm{M_{\odot}}$ and apply a filter of $M_{\mathrm{G}} < 12.1~\mathrm{mag}$ to ensure isochrone model accuracy. At the high-mass end, a horizontal cut at a fixed $M_{\mathrm{G}}$ would include metal-rich stars above $1.0~\mathrm{M_{\odot}}$, which undergo significant evolution. Instead, we use the iso-mass line for unevolved $1.0~\mathrm{M_{\odot}}$ stars, shown as the upper black dotted line in \Cref{fig:GCNS_cmd}. The resulting sample, colour-coded by density in \Cref{fig:GCNS_cmd}, comprises 233,217 stars for further analysis.

\section{Method} \label{sec:method}

\subsection{General description} \label{sec:general}

% \kp{Definitions in the article: 1) IMF: mass range, mass distribution for every component, single stars and components in binaries share the same IMF, averaged IMF over stellar populations in the solar neighbourhood; 2) binary fraction: system fraction, assume all multiple systems are binaries}

We develop a population synthesis model to reproduce the observational properties of field stars within the 100-pc solar vicinity, starting from a stellar initial mass function (stellar IMF). In this section, we first clarify key definitions, including those of the stellar IMF and binary fraction. We then identify critical corrections for observational biases and physical effects. Finally, we describe the equal-frequency binning strategy used throughout our analysis to minimize statistical biases.

To extract the initial mass function from field stars, several definitions must be clarified.
First, the IMF that we aim to determine in this article is the average stellar IMF over multi-metallicity populations in the solar neighbourhood.
The stellar IMF represents the initial mass distribution of a single star-formation event. Although in principle the mass function derived from field stars is defined as a ``composite IMF'' according to Kroupa's review, what we derive is close to the initial mass function because low-mass stars hardly evolve. Thus, our derived IMF can be treated as the averaged stellar IMF for low-mass stars in the solar neighbourhood.
The IMF we derive here represents the distribution of all stars (not systems), thereby counting all individual components in multiple systems \citep{kroupa2013}.
Second, the term ``binary fraction'' refers to the proportion of binary systems (including resolved and unresolved ones) relative to the total number of stellar systems. Specifically, when calculating this fraction, each binary star system is treated as a single unit, regardless of the number of stars it contains. 

% \kp{bias that need to be noted: overall description, simplification}

% \kp{eventually we plan to consider: Unresolved multiple stars, variation of stellar mass-luminosity relation for stars with different metallicity, Malmquist bias, and Lutz-Kelker bias}

% \kp{describe our solutions briefly}

We aim to reassess the IMF in the low-mass regime ($m/\mathrm{M_{\odot}}<1.0$) using the field star sample from the 100-pc solar neighbourhood. Direct star counting and luminosity-to-mass conversion using a single mass-luminosity relation (MLR) curve are not feasible due to several potential issues: (1) sample completeness; (2) Lutz-Kelker bias; (3) stellar evolution; (4) unresolved multiple systems; and (5) variations in the stellar mass-luminosity relation (MLR) due to chemical composition, age, and spin \citep[as concluded by][]{kroupa2013}. 

In this study, we disregard the effects of stellar spin and age distribution, as they have negligible impact on the low-mass IMF ($m/\mathrm{M_{\odot}}<1.0$). For low-mass stars ($<1~\mathrm{M_{\odot}}$), rotation plays a minimal role in their position on the colour-magnitude diagram (CMD), as these stars are not expected to be rapid rotators \citep{bastian2009}.

Regarding stellar evolution, we conducted tests using PARSEC 1.2s stellar tracks \citep{bressan2012,chen2014,chen2015,tang2014,marigo2017,Pastorelli2019,Pastorelli2020} assuming a constant star formation history from 1 to 10 Gyr. For stars with $4.1<M_{\mathrm{G}}<12.1~\mathrm{mag}$ and initial masses $M_{\mathrm{ini}}<1~\mathrm{M_{\odot}}$, only 1.1\% of evolved stars would be excluded by our sample cuts (the $1~\mathrm{M_{\odot}}$ iso-mass line and $M_{\mathrm{G}}=12.1~\mathrm{mag}$ cut). Among evolved stars that survive our selection criteria, only 0.7\% show mass differences between their present-day and initial masses. Therefore, $\alpha_2$ might be slightly biased toward smaller values due to missing a small fraction of evolved bright stars, but this effect is negligible. 

For pre-main sequence (PMS) stars, cross-matching our sample with \citet{zari2018} yields only 424 PMS stars (0.2\% of our sample after cuts). While the true PMS fraction may be higher given that catalogue prioritizes purity over completeness, the PMS contamination remains minimal. Since we adopt a three-part power-law IMF with separate treatment below $0.25~\mathrm{M_{\odot}}$ (see \Cref{sec:dis_metal_bias} for specific reason), and PMS effects are most significant at lower masses, our reported $\alpha_1$ and $\alpha_2$ values above $0.25~\mathrm{M_{\odot}}$ are minimally affected.

By excluding evolutionary effects from our model, we avoid making additional assumptions about the star formation history (SFH) of field stars.

Consequently, our method focuses on correcting biases from: (1) sample completeness; (2) Lutz-Kelker bias; (3) unresolved multiple systems; and (4) variations in the stellar mass-luminosity relation (MLR) due to metallicity.

We assessed the sample completeness using the $\mathrm{V/V_{max}}$ technique in \Cref{sec:GCNS}, revealing that our sample is volume-complete in the mass range under study. The Lutz-Kelker bias is a systematic bias whereby, on average, observed trigonometric parallaxes are larger than true trigonometric parallaxes \citep{lutz1973}. Thus, absolute magnitudes obtained from trigonometric parallaxes can also be biased. The extent of this bias depends on the relative error, $\sigma_{\varpi}/{\varpi}$. According to \citet{lutz1973}, when $\sigma_{\varpi}/{\varpi}$ increases from 0.025 to 0.175, the correction $<\Delta M>$ increases from -0.01 to -0.43. Only $1\%$ of our sample has $\sigma_{\varpi}/{\varpi}>0.05$, so the correction is insignificant. Consequently, in our model we need only consider the influence of the stellar mass-luminosity relation (MLR) varying with metallicity and unresolved multiple systems.

MLR varying with metallicity: The position of a star on the CMD (i.e., colour-magnitude diagram) can be determined by its mass, age, and metallicity. Generally, these three parameters are degenerated at some point, hence we cannot estimate stellar mass (also age, and metallicity) based solely on luminosity and colour. Fortunately, main-sequence stars on the CMD only slightly move with age, and their positions are largely determined by their masses, and their metallicities. Here, we plan to derive a metallicity distribution first based on \textsl{Gaia} XP spectra metallicity measurement \citep{li2023} and we will sample the metallicity for our model star according to this distribution in the forward modeling.

Unresolved multiple systems: We only consider the binaries instead of other multiple systems in the field population as other multiple systems only make up for about $10\%$ of all systems for solar-type stars according to \citet{moe2017}. 

The most straightforward way to model unresolved binaries is to pair two groups of stars drawn randomly from the same IMF, then synthesize the luminosity for the unresolved ones to match the observations. To ensure consistency with observed mass ratio distributions in field binaries \citep[e.g., large sample wide binary statistics produced by][]{raghavanSurveyStellarFamilies2010, el-badry2019a}, we incorporate an operator that acts on both the period distribution and mass-ratio distribution in our model, accounting for the evolutionary processes that modify our assumed initial binary population distributions to the observed ones in the solar neighbourhood. This approach was pioneered by \citet{kroupa1995b} for IMF measurements and has been further developed in subsequent works \citep{marks2011a, dabringhausen2022}.

Specifically, the overall shape of the observed mass-ratio distribution (slopes, break point, etc.) differs substantially from random pairing, and a feature of the present-day mass-ratio distribution called the twin phenomenon is identified, which manifests as a statistical excess of nearly equal-mass binaries ($0.95<q<1$). This discrepancy would lead to an inaccurate description of the binary main sequence in our forward model. Thus, we introduce a procedure to modulate the random-pairing-like mass-ratio distribution to the observed one. The most intuitive approach is to mimic the dynamical processes of binary systems in star clusters to effect this conversion. Dynamical evolution acts primarily as a function of the binding energy of the binary system. Disrupted binaries enter the field as single stars, thereby affecting the shape of the observed mass function. However, the mass-ratio distribution has been suggested to be insensitive to dynamical evolution in energy after formation \citep{parkerBinaryCompanionMass2013}, so we construct a dynamical operator as a function of both binding energy and mass ratio, which will be described in detail in \Cref{sec:procedure}.

\textit{Important feature of our population synthesis model:} We describe the binary population in the field with particular care, especially regarding their proportion and mass-ratio distribution. A key assumption of our dynamical evolution model is that stars in the solar neighbourhood were born in embedded clusters, after which most clusters were disrupted and their members mixed to become field stars \citep{kroupa1995b,marks2011a}. This mixing process refers specifically to the mixing of stellar density, metallicity, binary fractions, and binary orbital parameter distributions (period distribution, mass-ratio distribution, etc.). Therefore, what we deduce are the parameters of a typical star cluster whose properties after dynamical evolution can describe the present-day properties of field stars in the solar neighbourhood.

% \kp{mention our binning strategy}

\textit{Binning strategy:} For all binning operations in this article, we adopt the equal-frequency binning strategy proposed by \citet{maizapellaniz2005a}, which minimizes the bias caused by unbalanced Poisson error in each bin.

\subsection{Procedure} \label{sec:procedure}

\begin{table*}
  \caption{Nomenclature and symbols used throughout this study.}
  \label{tab:notation}
  \newcolumntype{L}[1]{>{\raggedright\arraybackslash}p{#1}}
  \newcommand{\indentpar}[1]{\noindent\hangindent=2em\hangafter=1 #1}
  \renewcommand{\arraystretch}{1.4}  % Adjust this value to change row spacing
  \begin{tabular}{cL{10cm}c}
    \hline
    Symbol & \multicolumn{1}{c}{Description} & Unit\\
    \hline
    $D$ & Distance from the Sun to a stellar system; & pc\\
    $m_1$, $m_2$ & Primary and secondary component masses in a binary system; & $\mathrm{M_{\odot}}$\\
    $P$ & Orbital period of a binary system; & day\\
    $\mathrm{[Fe/H]}$ & Stellar metallicity; & dex\\
    \tabincell{c}{$M_{\mathrm{G,1}}$, $M_{\mathrm{G_{BP},1}}$, $M_{\mathrm{G_{RP},1}}$ \\$M_{\mathrm{G,2}}$, $M_{\mathrm{G_{BP},2}}$, $M_{\mathrm{G_{RP},2}}$} & \indentpar{$G$, $G_{\mathrm{BP}}$, and $G_{\mathrm{RP}}$ absolute magnitudes for the primary and secondary components, respectively;} & mag\\
    $a_{\mathrm{phy}}$ & Physical separation between binary components; & AU\\
    $\theta$ & Projected angular separation on the sky; & arcsec\\
    $\mathrm{log_{10}}E_{\mathrm{b}}$ & Binary binding energy (logarithm); & $\mathrm{M_{\odot}~km^2~s^{-2}}$\\
    $q$ & \indentpar{Mass ratio, defined as $m_2/m_1$;} & /\\
    \tabincell{c}{$\mathrm{phot_{dis1}}$, $\mathrm{phot_{dis2}}$ \\ $\mathrm{phot_{rb1}}$, $\mathrm{phot_{rb2}}$ \\ $\mathrm{phot_{ub}}$} & \indentpar{Photometric properties ($M_{\mathrm{G}}$, $M_{\mathrm{G_{BP}}}$, $M_{\mathrm{G_{RP}}}$) for disrupted, resolved, and unresolved systems; subscripts ``rb1'', ``rb2'' denote resolved binary components, while ``dis1'', ``dis2'' denote disrupted binary components.} & mag\\
    $\alpha_0$, $\alpha_1$, $\alpha_2$, $m_{\mathrm{break}}$ & \indentpar{IMF power-law indices for the three segments (see \Cref{eq:6}) and the break mass;} & /\\
    $\gamma$, $\delta$, $\mathrm{loc}$, $\mathrm{scale}$ & \indentpar{Parameters of the Johnson's $S_U$ distribution describing metallicity;} & /\\
    $\mathrm{log_{10}}\rho$ & \indentpar{Characteristic density of the natal cluster in the dynamical evolution model;} & $\log_{10}(\mathrm{M_{\odot}~pc^{-3}})$\\
    $\gamma_1$, $\gamma_2$, $q_{\mathrm{break}}$ & \indentpar{Parameters of the present-day mass-ratio distribution, modeled as a piecewise power law;} & /\\
    $F_{\mathrm{twin}}$ & \indentpar{Twin binary fraction, defined in \Cref{eq:twin};} & /\\
    $N_{\mathrm{twin}}$, $N_{\mathrm{rgl}}$ & \indentpar{Number of twin binaries and number of regular binaries surviving dynamical evolution;} & count\\
    $res$ & \indentpar{Effective angular resolution of \textsl{Gaia} DR3 for the 100-pc sample;} & arcsec\\
    $noise$ & \indentpar{Standard deviation of Gaussian noise added to simulated $\Delta M_{\mathrm{G}}$ distributions.} & mag\\
    \hline
  \end{tabular}
\end{table*}

\begin{figure*}
    \centering
    \includegraphics[width=0.7\linewidth]{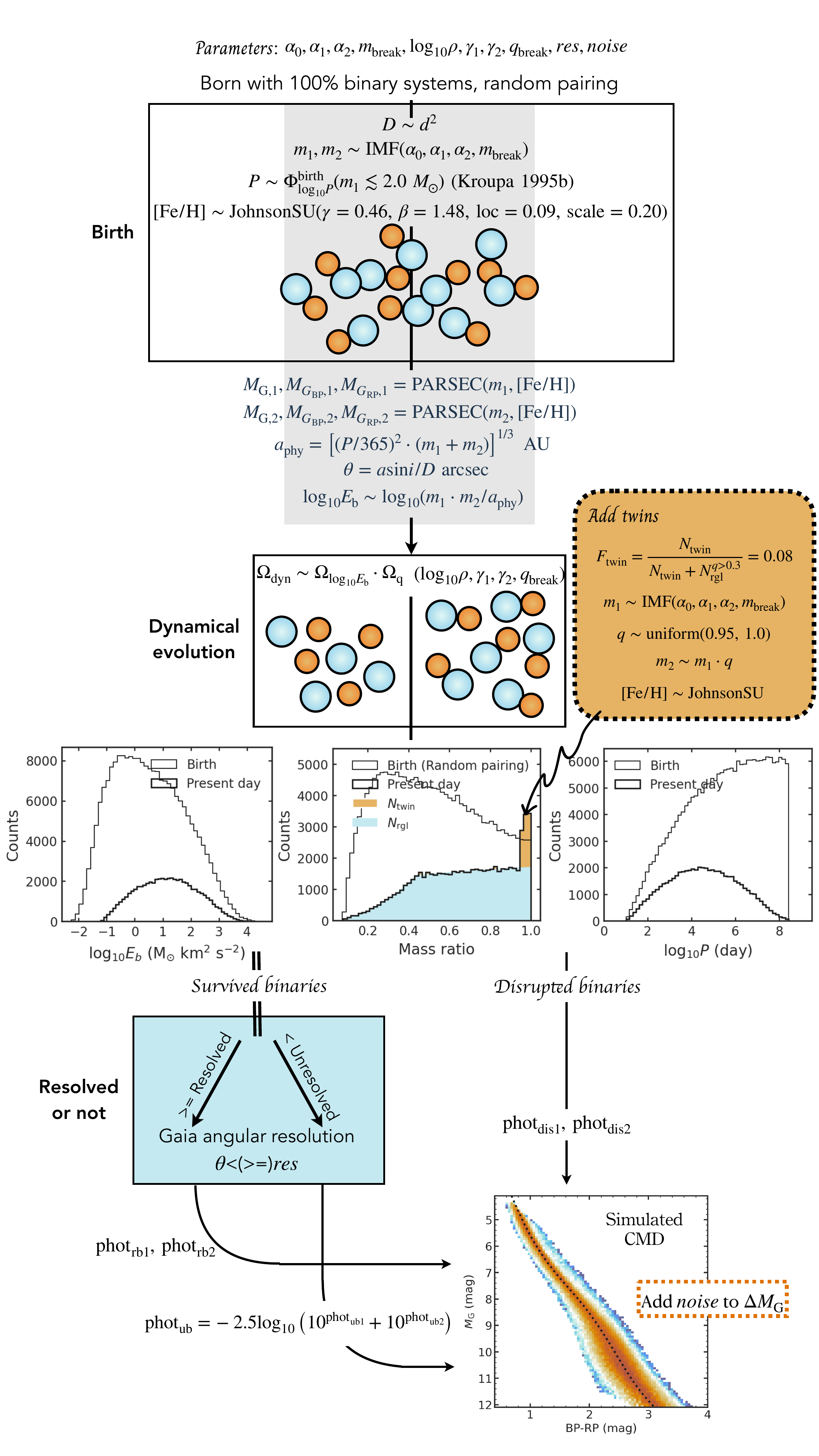}
    \caption{Schematic flow diagram of the population synthesis methodology employed in this study. The model assumes that all stars form in embedded clusters with an initial binary fraction of 100\%, where both components are sampled from the same IMF and paired randomly (uppermost panel). Subsequent cluster-mode dynamical evolution disrupts a subset of binary systems ($\mathrm{phot_{dis}}$), transforming the period and mass-ratio distributions to their present-day configurations. Surviving binaries are then classified according to \textsl{Gaia} DR3's angular resolution into resolved ($\mathrm{phot_{rb}}$) and unresolved ($\mathrm{phot_{ub}}$) systems. The photometric properties of unresolved binaries are recalculated to account for their combined flux. The notation $X \sim p(X)$ in this paper means that $X$ is drawn from the probability distribution $p(X)$.}
    \label{fig:model}
\end{figure*}

% \kp{how we generate our model stars in the solar neighbourhood}

We use the population synthesis method to generate the synthesized CMD for field stars in the 100-pc solar neighbourhood and constrain the IMF and binarity. In this population synthesis, we incorporate novel elements that we term ``cluster-mode'' dynamical evolution of binary orbital parameter distributions to fit binary properties in the solar neighbourhood.

In general, we assume that stars form in clusters with 100\% binaries, and that both components in a binary system are sampled from the same IMF and paired randomly at birth. Subsequently, the stars undergo cluster-mode dynamical evolution during which some binary systems disrupt, modifying the binary period and mass-ratio distributions to their present-day forms. Due to the angular resolution of \textsl{Gaia}, some of the surviving binary systems cannot be resolved. Their observed luminosity and colour shift relative to those of the primary stars (the brighter component). Thus, we determine which systems cannot be resolved and calculate their shifted luminosity and colour, which then constitute the binary main sequence. Finally, we compare the synthesized present-day CMD with the observed one to constrain the IMF and binary properties.

We adopt the following procedure specifically to generate one set of model particles:

I. \textit{Birth distribution}

(i). \textit{Initial mass function (IMF)}: Since synthesizing binary population by random pairing requires us to expand the mass range that we samples from, we choose the functional form for the initial mass function to be a three-part piece-wise power-law distribution. The parameters are $\alpha_0$ (nuisance parameter for isolating biased mass range $0.08<m\leq0.25~M_{\odot}$), $\alpha_1$ (for $0.25<m\leq m_{\mathrm{break}}$), $\alpha_2$ (for $m_{\mathrm{break}}<m\leq1.0~M_{\odot}$), and $m_{\mathrm{break}}$ to represent the power-law indices and the positions of the break point as in \Cref{eq:6}. 
We adopt this three-part formulation to account for potential bias introduced by our magnitude cut at $M_{\mathrm{G}} = 12.1$~mag, which preferentially excludes higher-metallicity low-mass stars that are fainter when obtaining the population-averaged stellar IMF (see discussion in \Cref{sec:dis_metal_bias}, showing that the $M_{\mathrm{G}}$-bias is negligible beyond $0.25~M_{\odot}$). 
We sample from the function for given parameters to be the mass collection of our model stars.

\begin{align}
\xi(m)=
\begin{cases}
C_0 \cdot m^{-\alpha_0}, & 0.08 \leq m/\mathrm{M_{\odot}} < 0.25,\\
C_1 \cdot m^{-\alpha_1}, & 0.25 \leq m/\mathrm{M_{\odot}} < m_{\mathrm{break}},\\
C_2 \cdot m^{-\alpha_2}, & m_{\mathrm{break}} \leq m/\mathrm{M_{\odot}} < 1.0
\end{cases}
\label{eq:6}
\end{align}

(ii). \textit{Metallicity distribution}: we sample from Johnson's SU distribution \citep{johnson1949} with $\gamma=0.91^{+0.06}_{-0.05}$, $\delta=1.61^{+0.05}_{-0.05}$, $\mathrm{loc}=0.10^{+0.01}_{-0.01}$ and $\mathrm{scale}=0.26^{+0.01}_{-0.01}$ to get the metallicity for every system. The derivation of these parameters is described in \Cref{sec:app_metal}. Since our dynamical evolution does not act on metallicity, it is reasonable not to distinguish birth metallicity distribution from present-day one.

(iii). \textit{Birth binary population}: we assume that the binary fraction is 100 per cent at birth \citep{kroupa1995b}. We sample two groups of masses from the initial mass function, and matched them together to form the binary systems. In this way, the component masses are paired randomly from the IMF, ensuring that both primaries and secondaries follow the same underlying mass distribution. This initial pairing, combined with subsequent evolution operator, reproduces the observed mass-ratio distribution in field binaries. Noted that the range in which we sample the mass should be wider than we need when conducting the random-pairing, so we choose to sample from $0.08~\mathrm{M_{\odot}}$ to $2.0~\mathrm{M_{\odot}}$.

(iv). \textit{Birth period distribution}: we adopt the birth period distribution obtained by \citet{kroupa1995b} for each binary system.

\begin{equation}
\Phi_{\mathrm{log_{10}}P}^{\mathrm{birth}}(m_1 \lesssim 2.0~\mathrm{M_{\odot}})=2.5\dfrac{\mathrm{log_{10}}P-1}{45+(\mathrm{log_{10}}P-1)^2},
\label{eq:8}
\end{equation}
where $P$ is the orbital period for binaries in the unit of day.

(v). \textit{Distance distribution}: we assume that the stellar volume density in the solar neighbourhood is constant, which means that the radial distance distribution satisfies $p(d) \propto d^2$. So we sample from this function to assign a distance to each system.

(vi). \textit{Projected separations}: With period and mass for each component in a system, we can calculate the physical separation $a_{\mathrm{phy}}$ of the binary in the unit of AU by Kepler's Third Law. We then adopt a random binary orientation in the inference to get the projected separation $a_{\mathrm{proj}}$. And we simply assume that binaries orbit in circular motion, so the projected distance $a_{\mathrm{proj}}$ will be $a\times \mathrm{sin}i$, where $i$ is the inclination of the orbit plane. The system's observed angular distance is consequently

\begin{equation}
\theta = \dfrac{a_{\mathrm{proj}}}{d}
\label{eq:9}
\end{equation}

~\\

II. \textit{Resolved or unresolved binary systems}

We determine whether a systems is resolved by comparing the angular resolution of \textsl{Gaia} and the projected angular distance of the system. According to \citet{collaborationGaiaEarlyData2021}, the incompleteness in close pairs of stars starts below about 1.5 arcsec, which can be regarded as effective angular resolution of \textsl{Gaia} EDR3. This value will change for different regions in the sky, so we turn it into a parameter as well in our model as $res$. The systems with projected angular distances greater than $res$ are considered resolved binary, while the others are unresolved binary. We need to synthesize the luminosity and colour of the two components for the unresolved systems.

~\\

III. \textit{Binary orbit evolution}: from birth distribution to present-day distribution

(i). \textit{Cluster-mode dynamical evolution operator}: evolution of binary period and binding energy distributions

\begin{figure}
    \centering
    \includegraphics[width=0.8\linewidth]{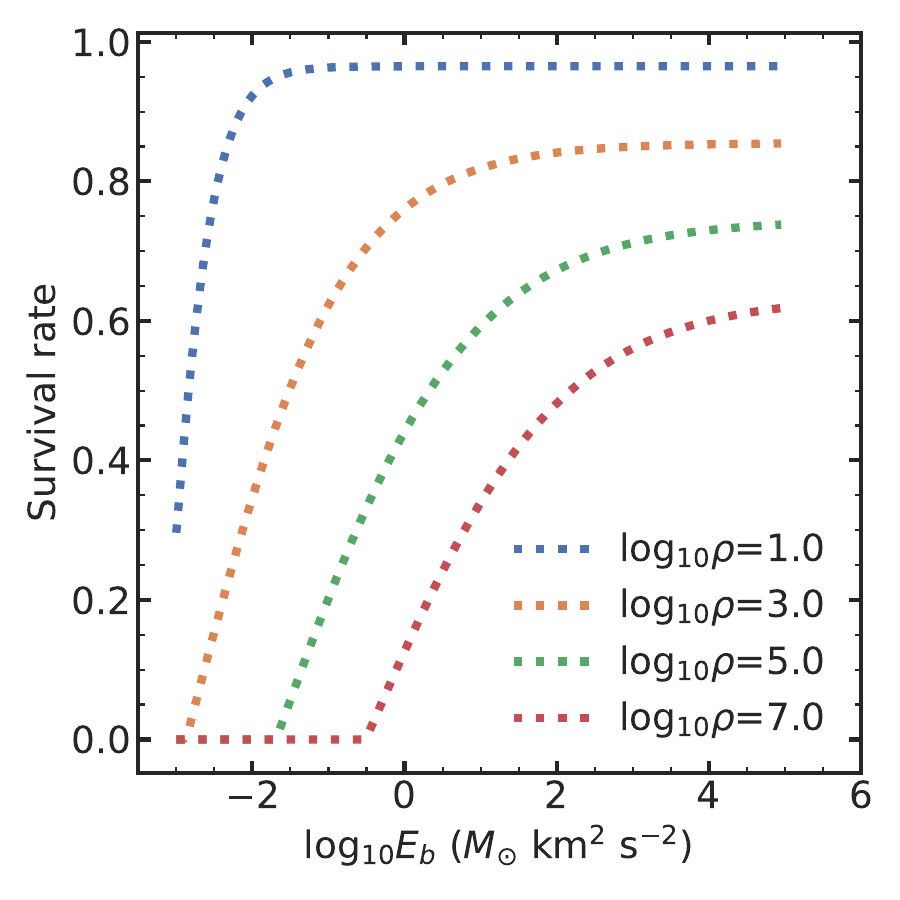}
    \caption{Dynamical evolution operator from \citet{marks2011a}. The function reflects the survival rate of binary systems under certain binding energy in the dynamical evolution. The plot shows the variation of the functional form when different characteristic cluster densities $\mathrm{log_{10}}(\rho/\mathrm{M_{\odot}~pc^{-3}})$ are set up.}
    \label{fig:dyna}
\end{figure}

In this paper, we will refer to the approach used by \citet{kroupa1995b} to describe the dynamical evolution of binary orbital parameters in the solar neighbourhood within a characteristic stellar cluster, known as a dominant mode cluster. We note that an alternative comprehensive approach is provided by the BIPOS1 code \citep{dabringhausen2022}, which models field binary populations by integrating over cluster properties. However, we adopt the dynamical evolution framework from \citet{marks2011a} as it allows us to treat the IMF slopes as free parameters, which is essential for our inference. The dynamical evolution of binaries in the cluster primarily acts on their binding energy $E_\mathrm{b}$, which is defined as \Cref{eq:10}.

\begin{equation}
E_\mathrm{b} = -\mathrm{G}\dfrac{m_1 m_2}{2a}
\label{eq:10}
\end{equation}
$\mathrm{G}$ is the gravitational constant, $m_1$ and $m_2$ refer to the component masses, and $a$ is the physical separation of the system.

Based on Kepler's Third Law, we can convert the right hand side of the \Cref{eq:10} to an equation relating to our sampling parameters in \Cref{sec:procedure} I.

\begin{equation}
\mathcal{\epsilon} = (2\pi \mathrm{G})^{2/3} m_{\mathrm{pr}}^{2/3} P^{-2/3} \dfrac{q}{(1+q)^{1/3}}
\label{eq:11}
\end{equation}
where $m_{\mathrm{pr}}$ is the mass of the primary star (heavier one), $P$ is the period, and $q$ is the mass ratio of the system.

% \kp{only consider the disruption of binary in the cluster, ignore other process like capture (takes up very small proportion), move to discussion!}

This paper focuses solely on the process of binary dissolution during dynamical evolution. \citet{marks2011a}, based on N-body simulations, appears to be very helpful since they provided empirical relationships for the distribution of binary orbital parameters. Specifically, they derived the survival probabilities of binary systems ($\Omega_{\mathrm{log_{10}}E_{\mathrm{b}}}$) at different evolutionary times and at different binding energies in \Cref{eq:12} for a given set of cluster parameters, primarily the average density of the cluster (denoted as the parameter $\mathrm{log_{10}}(\rho/\mathrm{M_{\odot}~pc^{-3}})$ afterwards).

\begin{align}
\Omega_{\mathrm{log_{10}}E_{\mathrm{b}}}(\mathcal{\epsilon}, t) = 
\begin{cases}
\dfrac{\mathcal{A}}
{1+\mathrm{exp\left[\mathcal{S}(\mathcal{\epsilon}-\mathcal{\epsilon}_{\mathrm{cut}})\right]}}-\dfrac{\mathcal{A}}{2} \quad \mathcal{\epsilon} \geq \mathcal{\epsilon}_{\mathrm{cut}} \\
0 \quad \mathrm{otherwise}
\end{cases}
\label{eq:12}
\end{align}
where $\mathcal{\epsilon}$ represents the binding energy of the system, while $\mathcal{A}$, $\mathcal{S}$ and $\mathcal{\epsilon}_{\mathrm{cut}}$ are parameters that are functions of time $t$ and the characteristic density of the stellar cluster $\mathrm{log_{10}}(\rho/\mathrm{M_{\odot}~pc^{-3}})$. Visualized form of this function can be seen in \Cref{fig:dyna}. 

% \kp{when does the evolution cease, why}

According to \citet{marks2011a}, the evolution of binary orbital parameter distribution during the dynamical evolution of star clusters ceases after approximately 5 million years (Myr) of formation. 
%\kp{This is attributed to the expansion of the star cluster during this period, resulting in a decrease in the cluster's gravitational field, making it more challenging for the surviving binaries to be affected. Eventually, the system reaches equilibrium.} 
Therefore, in our model, we only apply the dynamical evolution operator at $t=5~\mathrm{Myr}$, $\mathcal{A}$, $\mathcal{S}$ and $\mathcal{\epsilon}_{\mathrm{cut}}$ are now only functions of the characteristic density of the stellar cluster $\mathrm{log_{10}}(\rho/\mathrm{M_{\odot}~pc^{-3}})$ as in \Cref{eq:13,eq:14,eq:15}. $\mathrm{log_{10}}(\rho/\mathrm{M_{\odot}~pc^{-3}})$ is then treated as fitting parameters in our model. The information on the observed present-day binary fraction in the data can constrain $\mathrm{log_{10}}(\rho/\mathrm{M_{\odot}~pc^{-3}})$, as a higher characteristic cluster density implies a lower survival probabilities $\Omega_{\mathrm{log_{10}}E_{\mathrm{b}}}$, thus a larger number of disrupted binaries, leading to a smaller binary fraction at present day.

\begin{figure}
    \centering
    \includegraphics[width=0.8\linewidth]{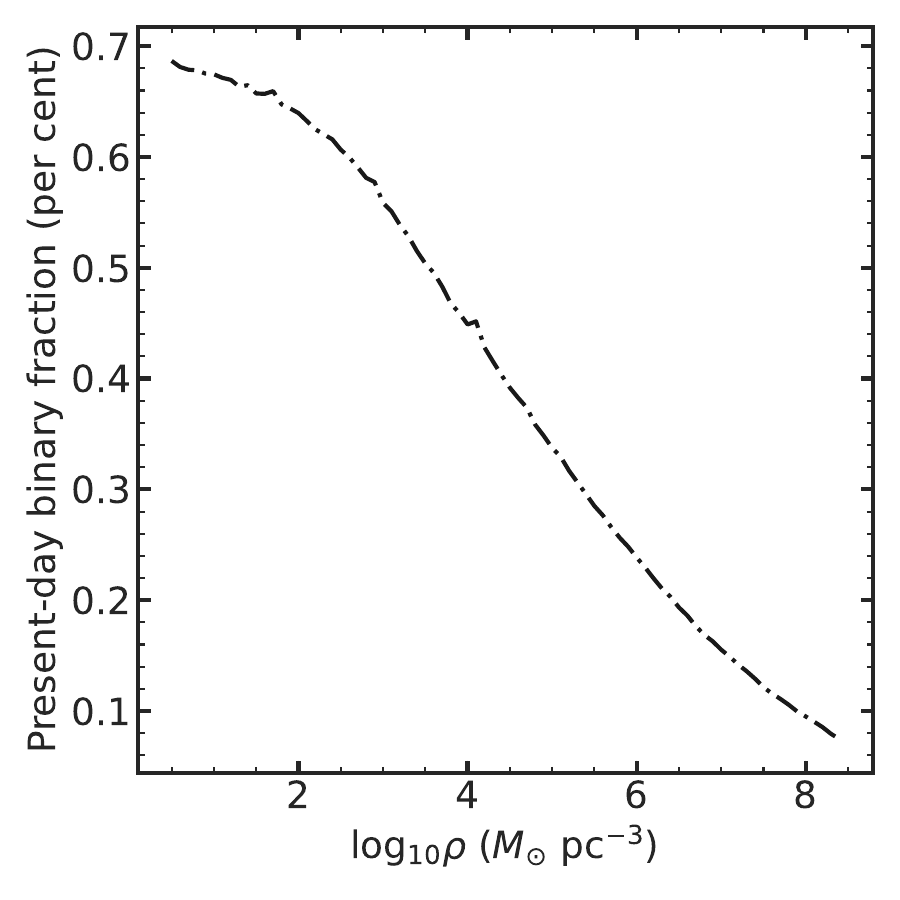}
    \caption{Relation of the characteristic cluster density in our dynamical evolution operator and subsequent present-day binary fraction.}
    \label{fig:rho_fb}
\end{figure}

\begin{align}
\mathcal{\epsilon}_{\mathrm{cut}}=
\begin{cases}
~-4.40+0.54 \cdot \mathrm{log_{10}}\rho \quad \mathrm{if~result}>3.2 \\
~-3.2 \quad \mathrm{otherwise}
\end{cases}
\label{eq:13}   
\end{align}

\begin{align}
\mathcal{A}=
\begin{cases}
~1.97-0.10 \cdot \mathrm{log_{10}}\rho \quad \mathrm{if~result}\leq 2 \\
~2 \quad \mathrm{otherwise}
\end{cases}
\label{eq:14}   
\end{align}

\begin{equation}
\mathcal{S}=-\dfrac{1}{\mathrm{exp}\left[1.47 \cdot (\mathrm{log_{10}}\rho-1.35)\right]}-0.82,
\label{eq:15}   
\end{equation}
when $t=5~\mathrm{Myr}$ in \citet{marks2011a}.

% \kp{mass-ratio distribution unchanged}

The rightmost one of the middle plots in \Cref{fig:model} illustrate the changes in the distribution of binary periods after the application of the dynamical evolution operator under certain parameter settings. We find that when we only apply \Cref{eq:12} to the binary population, the shape of the period distribution is significantly altered, while the shape of the mass-ratio distribution remains unchanged. This finding is consistent with the statement made by \citet{parkerBinaryCompanionMass2013} that the distribution of mass ratios is insensitive to dynamical evolution. It suggests that binaries with the same mass ratio experience similar dissolution rates during this process.

(ii). \textit{Mass-ratio distribution evolution operator}

To achieve the transformation of the mass-ratio distribution from a random-pairing form to the present-day distribution, which can be constrained using observational data, we describe the present-day mass-ratio distribution using two power-law segments and a break point. The parameters involved are denoted as $\gamma_1$, $\gamma_2$, and $q_{\mathrm{break}}$. The random-pairing mass-ratio distribution is dependent only on the form of the IMF. Hence, we can derive an evolution operator for the mass ratio, denoted as $\Omega_{\mathrm{q}}$, by dividing the present-day mass-ratio distribution by the mass-ratio distribution obtained from random pairing with a given IMF. The overall expression of the operator can be written as follows:

\begin{equation}
\Omega=C \cdot \Omega_{\mathrm{log_{10}}E_{\mathrm{b}}} \cdot \Omega_{\mathrm{q}}
\label{eq:16}   
\end{equation}
The operator is normalized by C to ensure that the overall dissolution rate of the binary systems remains the same as when only the dynamical evolution operator $\Omega_{\mathrm{log_{10}}E_{\mathrm{b}}}$ is applied.

~\\

IV. \textit{Twin binaries}

In the history of multiple system studies, plenty of work has reported the discovery of a puzzling feature called twin phenomenon \citep{raghavanSurveyStellarFamilies2010, duchene2013, moe2017, el-badry2019a}, which can also be located in our wide binary mass-ratio distribution at $q=0.95-1.0$ in \Cref{fig:wb_q} with $F_{\mathrm{twin}}=0.08$. A peak at $q \approx 1$ can arise naturally in the dynamical population synthesis of \citet{marks2011a} through their pre-main sequence eigenevolution, as illustrated in fig.~1 of \citet{kroupa2025}. However, that feature is produced only for short-period binaries, whereas our wide-binary sample still shows an excess at $q \approx 1$ (\Cref{fig:wb_q}) and these need to be added as extra systems into the modelling. \citet{el-badry2019a} suggests that twins likely form from a mechanism distinct from other binaries, such as competitive accretion from a circumbinary disc. Since twin binaries form through a different physical process and their mass ratios are close to unity, meaning both components can be assumed to independently follow the stellar IMF, it is physically justified to add twin binaries to our synthesis model as an additional population after the cluster dynamical evolution process rather than subjecting them to the same dynamical modulation applied to regular binaries.

When regular binaries (form from non-twin mechanism) finish their dynamical process, we can then calculate how many twins should we add to our model, since our definition of $F_{\mathrm{twin}}$ is 

\begin{equation}
F_{\mathrm{twin}} = \dfrac{N_{\mathrm{twin}}}{N_{\mathrm{twin}}+N_{\mathrm{rgl}}^{q>0.3}}
\label{eq:twin}   
\end{equation}
where $N_{\mathrm{rgl}}^{q>0.3}$ means binaries forming in regular process with mass ratio larger than 0.3. In our model, \textit{regular process} means that they are sampled from the IMF and randomly paired at birth, and then go through our dynamical evolution operator. This definition is the same as \citet{moe2017} and \citet{el-badry2019a}. 

There will be resolved and unresolved twins as well, and we need their present-day period distribution to realize the process in our model. Here we simply assume that twins share the same period distribution as regular binaries. Therefore, we respectively calculate the number of resolved and unsolved twin binary systems $N_{\mathrm{twin,rb}}$, $N_{\mathrm{twin,ub}}$ with \Cref{eq:twin} by substituting $N_{\mathrm{rgl}}^{q>0.3}$ with $N_{\mathrm{rgl,rb}}^{q>0.3}$ and $N_{\mathrm{rgl,ub}}^{q>0.3}$.

With the number of resolved and unresolved twins, we can then sample their mass ratio as uniform distribution from 0.95 to 1.0, and sample the masses of the primary with the IMF defined in \Cref{eq:6}. The masses of secondaries are therefore available. We apply the same metallicity distribution as regular binaries to these twins. And through PARSEC model, the photometry of each model star can be acquired interpolating on the grid of mass and metallicity. For unresolved twins, we should synthesize the photometry of two components.

~\\

V. \textit{Output of the model}

\begin{figure*}
    \centering
    \includegraphics[width=0.8\linewidth]{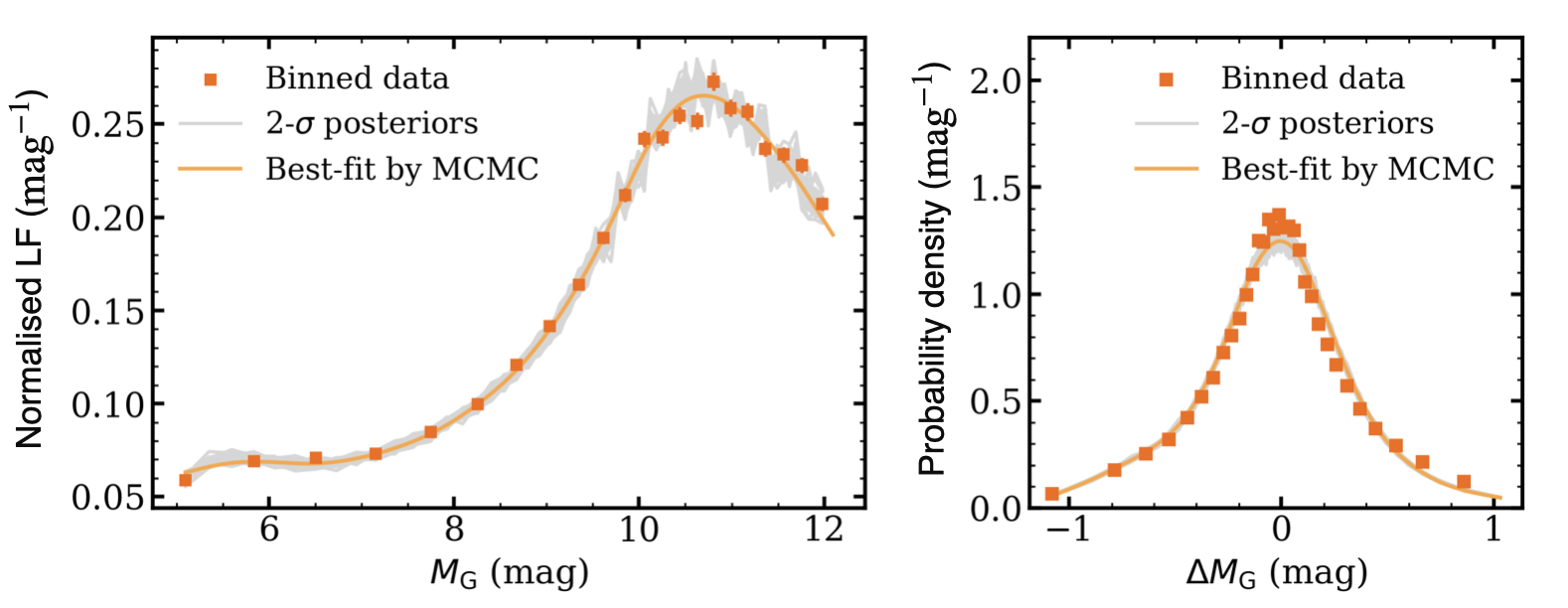}
    \caption{Comparison between observed luminosity function $\phi_{\mathrm{obs}}(M_{\mathrm{G}})$, $\Delta M_{\mathrm{G,\mathrm{obs}}}$ distribution and our best-fit model. Both distributions are normalized to unit area, so the y-axis gives probability density rather than number density.}
    \label{fig:output}
\end{figure*}

% \kp{with PARSEC model, obtain the photometry for our model stars}

For all resolved binary systems obtained from the aforementioned procedure, as well as unresolved but disrupted binary systems, we directly utilize the PARSEC model version 1.2s to obtain corresponding values of $M_{\mathrm{G}}$, $M_{\mathrm{G_{BP}}}$, and $M_{\mathrm{G_{RP}}}$ through interpolation of sampled masses and metallicities. For unresolved yet surviving binary systems, we synthesize the system's luminosity and magnitude (via flux) after obtaining the brightness and colours of individual stars. We employ the PARSEC model with initial masses $M_{\mathrm{ini}}<1~\mathrm{M_{\odot}}$, metallicities $-1.0<\mathrm{[M/H]}<0.6$, and age $\mathrm{log_{10}} \mathrm{Age}=9.5$.

% \kp{fit for the ridge lines of the model CMD and oberved CMD, build the likelihood function on MG and dMG distribution.}

Finally, we obtain a set of model stars that incorporate the effects of unresolved binary systems. To mitigate the inaccuracies of stellar models at the low-mass end \citep{chen2014}, we separately fit the ridge lines $(\mathrm{BPRP}-M_{\mathrm{G}})_{\mathrm{c,mod}}$ and $(\mathrm{BPRP}-M_{\mathrm{G}})_{\mathrm{c,obs}}$ for the simulated CMD and the observed CMD, respectively, and subtract the ridge lines from the CMDs to get $\Delta M_{\mathrm{G}}$. After this, we consider the observational uncertainty by adding a Gaussian noise model to the simulated $\Delta M_{\mathrm{G}}$, where the means are set to the original simulated $\Delta M_{\mathrm{G}}$ value, and the variation $noise$ is assumed to be universal and set as a free parameter in our model. The observational uncertainty of $M_{\mathrm{G}}$ is insignificant relative to the span of the luminosity function.

Subsequently, by comparing the distributions of $M_{\mathrm{G}}$ and the noised residual distribution $\Delta M_{\mathrm{G}}$ after subtracting the ridge line, we can constrain the parameters $\alpha_0$, $\alpha_1$, $\alpha_2$, $m_\mathrm{break}$ (IMF), $\gamma_1$, $\gamma_2$, $q_{\mathrm{break}}$ (mass-ratio distribution), $res$ (average resolution of \textsl{Gaia}), $\mathrm{log_{10}}(\rho/\mathrm{M_{\odot}~pc^{-3}})$ (cluster-mode dynamical evolution), and $noise$. The whole process is depicted in the flowchart \Cref{fig:model}.

~\\

VI. \textit{Bayesian estimation of the best parameters}

% \kp{posterior}
% \kp{explain why we cut MG~(5.1, 12.1) (PARSEC model), dMG~(-1,1)}

We establish the likelihood function on the distribution of $M_\mathrm{G}$ (i.e. the luminosity function) and the distribution of $\Delta M_\mathrm{G}$ separately. The $\Delta M_\mathrm{G}$ is defined as 

\begin{equation}
\Delta M_{\mathrm{G}}=M_{\mathrm{G}}-M_{\mathrm{G,ridge}}(G_{\mathrm{BP}}-G_{\mathrm{RP}})
\label{eq:17}   
\end{equation}
which means the variation of $M_\mathrm{G}$ relation to the ridge line of the CMD. The information of binary fraction and mass-ratio can be represented better in $\Delta M_\mathrm{G}$ distribution, while the information of stellar mass distribution in $M_\mathrm{G}$ distribution.

The observational and model $M_\mathrm{G}$ and $\Delta M_\mathrm{G}$ data are binned in the same equal-binning strategy and the distributions can be denoted as $\left\{\phi_{\mathrm{obs,i}}\right\}$, $\left\{\phi_{\mathrm{mod,i}}\right\}$ and $\left\{\psi_{\mathrm{obs,i}}\right\}$, $\left\{\psi_{\mathrm{mod,i}}\right\}$, respectively. The likelihood function for one bin can be depicted as Gaussian distribution considering the observational uncertainties and Poisson error. The likelihood function and posterior are written in \Cref{eq:18} and \Cref{eq:19}.

\begin{align}
\mathcal{L}_i =
-\dfrac{1}{\sqrt{2\pi (\sigma^2_{\mathrm{obs,i}}+\sigma^2_{\mathrm{Poi,\phi_i}})}} \mathrm{exp} \left\{ -\dfrac{(\phi_{\mathrm{obs,i}}-\phi_{\mathrm{mod,i}})^2}{2 (\sigma^2_{\mathrm{obs,i}}+\sigma^2_{\mathrm{Poi,\phi_i}})}\right\}
\notag
\\-\dfrac{1}{\sqrt{2\pi (\sigma^2_{\mathrm{obs,i}}+\sigma^2_{\mathrm{Poi,\psi_i}})}} \mathrm{exp} \left\{ -\dfrac{(\psi_{\mathrm{obs,i}}-\psi_{\mathrm{mod,i}})^2}{2 (\sigma^2_{\mathrm{obs,i}}+\sigma^2_{\mathrm{Poi,\psi_i}})}\right\}
\label{eq:18}
\end{align}

\begin{align}
P(\alpha_0, \alpha_1, \alpha_2, m_{\mathrm{break}}, \gamma_1, \gamma_2, q_{\mathrm{break}}, \mathrm{log_{10}}\rho, {\rm res} \mid \left\{\phi_{\mathrm{obs,i}}\right\},\left\{\psi_{\mathrm{obs,i}}\right\})
\notag
\\=\texttt{prior}(\alpha_0, \alpha_1, \alpha_2, m_{\mathrm{break}}, \gamma_1, \gamma_2, q_{\mathrm{break}}, \mathrm{log_{10}}\rho, {\rm res}) \cdot \prod_i{\mathcal{L}_i}
\label{eq:19}
\end{align}

\subsection{Sample from the posterior} \label{sec:sample}
~\\

\begin{figure*}
    \centering
    \includegraphics[width=0.8\linewidth]{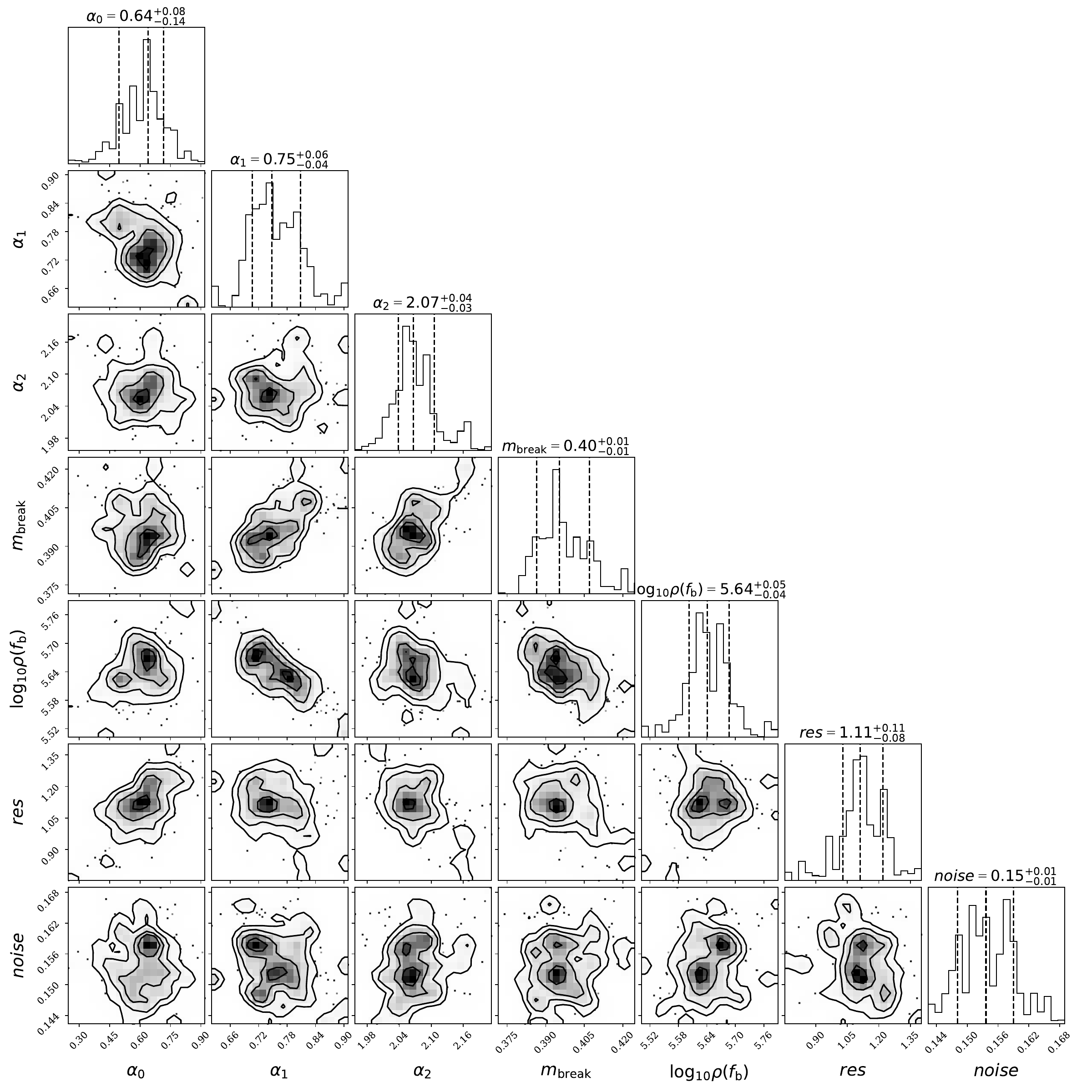}
    \caption{The posterior distribution of the model parameters with our main sample described in \Cref{fig:GCNS_cmd}.}
    \label{fig:mcmc}
\end{figure*}

% \kp{how we sample from the posterior}

% \kp{by grid-test, we find that the parameters of the mass-ratio distribution cannot be constrained well, give a very sharp gaussian prior at gamma1=0, gamma2=0, qbreak=0.5}

% \kp{auto-correlation time: a, moves, chains}

We have a high-dimensional posterior depicting the joint probability distribution of $\alpha_0$, $\alpha_1$, $\alpha_2$, $m_{\mathrm{break}}$, $\gamma_1$, $\gamma_2$, $q_{\mathrm{break}}$, $\mathrm{log_{10}}(\rho/\mathrm{M_{\odot}~pc^{-3}})$, and $res$. For a global view of the posterior distribution, we first draw a rough eight-dimensional grid for the posterior to inspect the parameter values near the maximum posterior. Through this process, we find that the parameters of the mass-ratio distribution cannot be constrained well, which we will further discuss in \Cref{sec:dis_dyn}. Therefore, we give a very sharp Gaussian prior for $\gamma_1 \sim \mathcal{N}(1.89, 0.23^2)$\footnote{We use the notation $N(\mu, \sigma^2)$ to denote a Gaussian distribution with mean $\mu$ and variance $\sigma^2$ in this paper. The notation $X \sim p(X)$ in this paper means that $X$ is drawn from the probability distribution $p(X)$.}, $\gamma_2 \sim \mathcal{N}(0.20, 0.13^2)$ and $q_{\mathrm{break}} \sim \mathcal{N}(0.44, 0.02^2)$ by the knowledge that we obtain from wide binaries, see \Cref{sec:app_q} for detailed calculation. Additionally, we add the prior for the binary fraction according to \citet{moe2017} to be $\sim \mathcal{N}(0.4, 0.05^2)$.

\Cref{tab:priors} summarizes all the priors adopted in our Bayesian inference. The Gaussian priors for the mass-ratio distribution parameters ($\gamma_1$, $\gamma_2$, $q_{\mathrm{break}}$) and the binary fraction are informed by wide binary statistics, while uniform priors are used for the IMF parameters and other model parameters to remain agnostic about their values.

\begin{table}
\centering
\caption{Prior distributions for all model parameters in the Bayesian inference.}
\label{tab:priors}
\begin{tabular}{lcc}
\hline
Parameter & Description & Prior Distribution \\
\hline
$\alpha_0$ & IMF slope ($0.08$--$0.25~\mathrm{M_{\odot}}$) & Uniform \\
$\alpha_1$ & IMF slope ($0.25$--$m_{\mathrm{break}}~\mathrm{M_{\odot}}$) & Uniform \\
$\alpha_2$ & IMF slope ($m_{\mathrm{break}}$--$1.0~\mathrm{M_{\odot}}$) & Uniform \\
$m_{\mathrm{break}}$ & IMF break point ($\mathrm{M_{\odot}}$) & Uniform \\
$\gamma_1$ & Mass-ratio slope ($q_{\mathrm{break}}$--$0.95$) & $\mathcal{N}(1.89, 0.23^2)$ \\
$\gamma_2$ & Mass-ratio slope ($0.2$--$q_{\mathrm{break}}$) & $\mathcal{N}(0.20, 0.13^2)$ \\
$q_{\mathrm{break}}$ & Mass-ratio break point & $\mathcal{N}(0.44, 0.02^2)$ \\
$f_{\mathrm{bin}}$ & Binary fraction (system) & $\mathcal{N}(0.4, 0.05^2)$ \\
$\mathrm{log_{10}}(\rho/\mathrm{M_{\odot}~pc^{-3}})$ & Cluster density & Uniform$^{\mathrm{a}}$ \\
$res$ & \textsl{Gaia} angular resolution (arcsec) & Uniform \\
\hline
\multicolumn{3}{l}{$^{\mathrm{a}}$The binary fraction prior constrains $\mathrm{log_{10}}(\rho/\mathrm{M_{\odot}~pc^{-3}})$ through} \\
\multicolumn{3}{l}{the dynamical evolution model (\Cref{sec:dis_dyn}).} \\
\end{tabular}
\end{table}

When running the MCMC program, we find that the randomness caused by sampling potentially leads to multi-modal posterior distribution. This feature makes \texttt{emcee} very hard to accomplish its required auto-correlation time and reach the point of convergence. Hence, we set the stretch scale parameter in \texttt{emcee} $a=1.0$ to increase the step size to prevent the walker from being stuck in local minimum. And we combine the \texttt{moves.DEMove} and \texttt{moves.DESnookerMove} as recommended by the package contributor, which is also helpful in multi-modal situations \citep{foreman-mackey2013}. We run the MCMC with 128 chains and 5,000 steps for each chain, and the final result is plotted in \Cref{fig:mcmc} in which the posterior distributions have been smoothed.

\section{Results} \label{sec:res}
~\\

\begin{figure}
    \centering
    \includegraphics[width=0.85\linewidth]{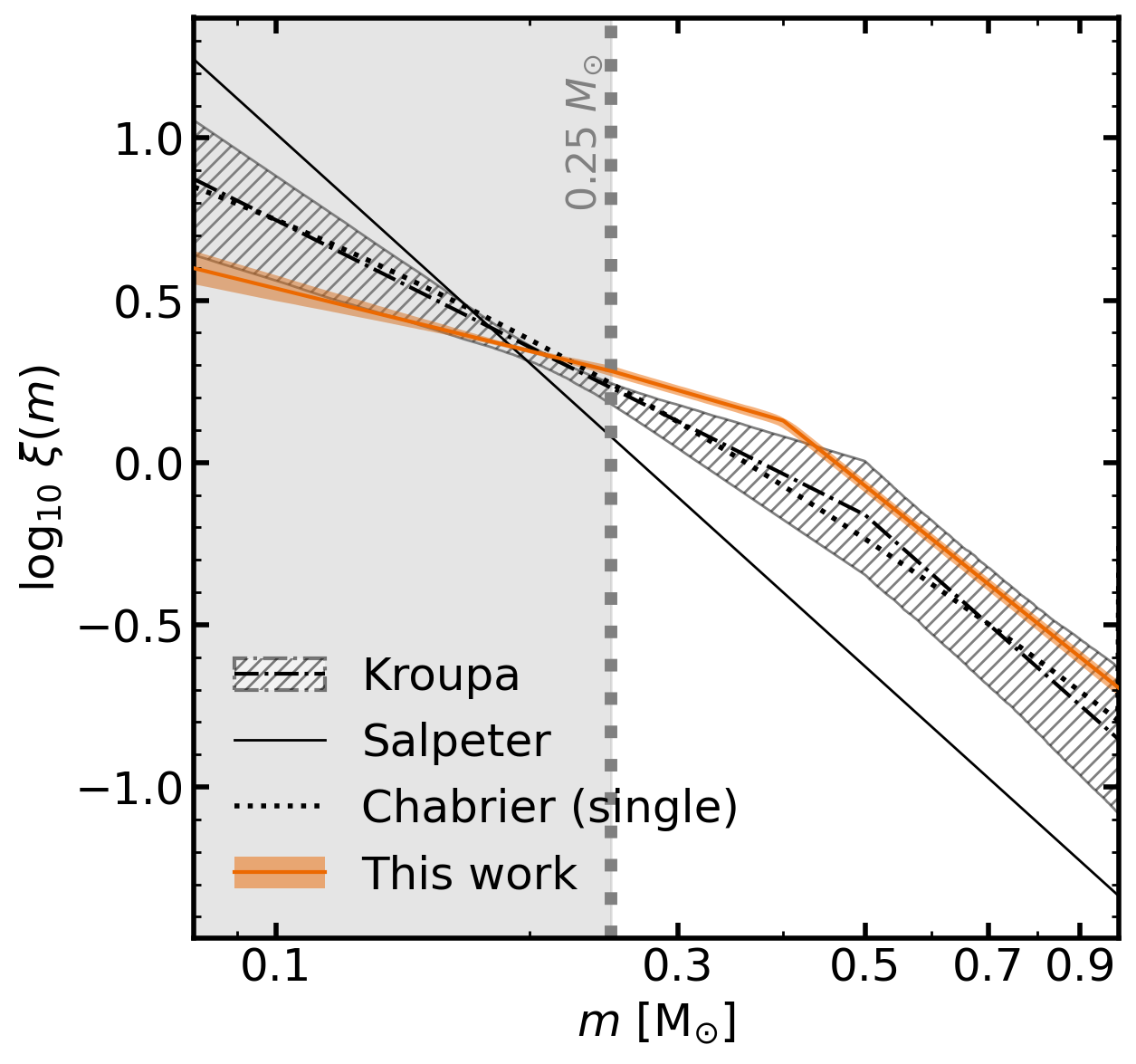}
    \caption{Comparison of the stellar IMF derived from \textsl{Gaia} DR3 data and our population synthesis model with canonical IMFs proposed by \citet{salpeter1955}, \citet{kroupa1993} and \citet{chabrierGalacticDiskMass2003}. 
    The black solid line denotes Salpeter's IMF with $\alpha=2.35$.  
    The black dash-dotted line and shaded area describe the Kroupa's IMF with $\alpha_1=1.3\pm0.5$ and $\alpha_2=2.3\pm0.3$. 
    The dotted line stands for Chabrier's IMF 
    The orange dotted line and shaded area defines the IMF derived in this work, where the gray shaded area masks out the biased range below $0.25~\mathrm{M_{\odot}}$ that should not be considered.
    All IMFs are normalized by the area under the curve from $0.08~\mathrm{M_{\odot}}$ to $1.0~\mathrm{M_{\odot}}$ in linear space.}
    \label{fig:IMF}
\end{figure}

% \kp{1. IMF: compare with Kroupa's, Salpeter's, Chabrier's result, correlation}

The posterior distribution is plotted in \Cref{fig:mcmc}, and the best estimation of the IMF is drawn in \Cref{fig:IMF} along with all the canonical IMFs. We adopt a three-part piece-wise power-law IMF parametrization (\Cref{eq:6}) to mitigate potential selection bias at the lowest masses (see \Cref{sec:dis_metal_bias} for detailed discussion). Our focus is on the well-constrained slopes for masses above $0.25~\mathrm{M_{\odot}}$: our best estimation for the averaged field star IMF is $\alpha_1=0.75^{+0.06}_{-0.04}$, $\alpha_2=2.07^{+0.04}_{-0.03}$ and $m_{\mathrm{break}}=0.40^{+0.01}_{-0.01}$, which is consistent with the Kroupa IMF but with much smaller uncertainties, owe to the unprecedented number and quality of the \textsl{Gaia} data.
%As Chabrier's IMF was confirmed consistent with Kroupa's, so our IMF is also indistinguishable from Chabrier's \citep{chabrierGalacticDiskMass2003}, but with well-defined uncertainties for each parameter.
Compared to Salpeter's IMF, our IMF (also Kroupa's and Chabrier's) are flatter for the mass range $m<0.40~\mathrm{M_{\odot}}$, avoiding the number divergence at the low-mass end. 
Additionally, our posterior distribution of break point shows slight correlation with $\alpha_1$ and $\alpha_2$ respectively.

Our IMF is averaged stellar initial mass function for all the components in the single star and binary systems in the solar neighbourhood field. The evolution of binary orbit evolution is taken considered by the way assuming that stars are born in binaries with the IMF we define and randomly-paired and dynamically evolved based on their binding energy and mass ratio.

% \kp{2. binary fraction}

The parameter $\mathrm{log_{10}}(\rho/\mathrm{M_{\odot}~pc^{-3}})$ in our model represents the typical density of the characteristic cluster. It will also determine the survived binary faction and can be compared with the present-day binary fraction in the observational data. The best estimation of $\mathrm{log_{10}}(\rho/\mathrm{M_{\odot}~pc^{-3}})$ is $5.64^{+0.05}_{-0.04}$. \Cref{fig:rho_fb} shows the relation between $\mathrm{log_{10}}(\rho/\mathrm{M_{\odot}~pc^{-3}})$ and present-day binary fraction, so our result reveals that the average binary fraction in the solar neighbourhood is approximately $26\%$.

% \kp{4. Gaia resolution}
The last parameter is the averaged angular resolution of \textsl{Gaia} for the 100-pc stars in the solar vicinity, and it is calculated as $1.11^{+0.11}_{-0.08}$ arcsec in our model. In the procedure, we distinguish the resolved and unresolved binaries by comparing the angular distance of the systems and this resolution. The official document of \textsl{Gaia} has stated that the effective resolution is 1.5 arcsec, at which the number counts of source pairs as a function of their separation drop below the expected line for a random source distribution for \textsl{Gaia} DR3 \citep{collaborationGaiaEarlyData2021}. This value can be affected by the region and source we are observing, and our determination of the resolution of \textsl{Gaia} is relatively reasonable, which indicates the reliability of our process.

\section{Discussion} \label{sec:dis}
~\\

\subsection{Dynamical evolution operator} \label{sec:dis_dyn}
% \kp{mass-ratio distribution, why we need to add a empirical function ourselves, why we can not constrain the gammas}

First, we clarify several points regarding our cluster-mode dynamical evolution operator. 

In this operator, we consider only the disruption of binaries in the cluster. According to the Heggie-Hills law \citep{heggie1975, hills1975}, hard binaries (those with binding energy greater than the average kinetic energy of the stellar cluster) tend to become even ``harder'' during dynamical evolution, while soft binaries (those with binding energy smaller than the average kinetic energy) become ``softer''. As a result, soft binaries eventually tend to dissolve. This paper focuses solely on the process of binary dissolution during dynamical evolution, whereas other processes such as binary capture, mergers, and binary coevolution have relatively weaker effects \citep{marks2011a}. We have also not considered changes in orbital parameters during this process, as \citet{kroupaburkert2001} showed that, compared to the broadening of binary separations, dissolution plays the dominant role in dynamical evolution. 

\begin{figure}
    \centering
    \includegraphics[width=0.85\linewidth]{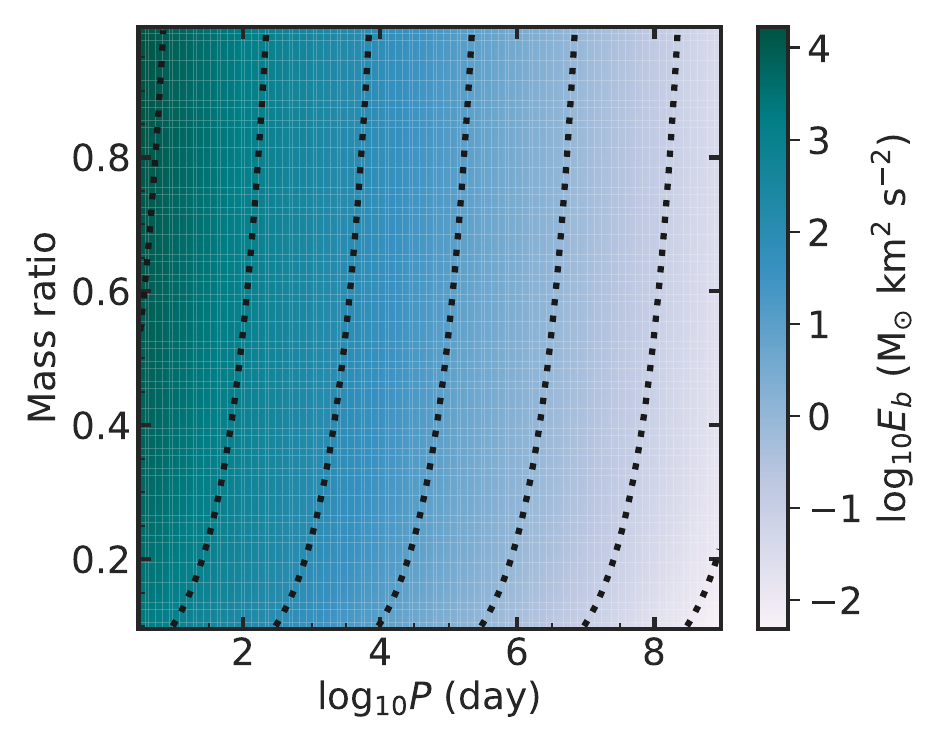}
    \caption{The influence of period and mass ratio on the binding energy according to Kepler's Third Law. We fix the mass of the primary star to $1.0~\mathrm{M_{\odot}}$ in this plot to demonstrate how the binding energy varies with period and mass ratio. The dotted lines depict the contour of the binding energy values. The binding energy mostly relies on the period.}
    \label{fig:p_q_lgeb}
\end{figure}

\textit{Why we need the additional part on the operator to modulate the mass ratio?} 
The initial idea of this work is to modify the mass-ratio distribution from the born random-pairing shape to today observed shape by a binary disruption process similar to what happens in a cluster's dynamical evolution. 
Physically, the dynamical evolution in a star cluster mainly relies on the binding energy of a binary system, and this is exactly how \citet{marks2011a} built their empirical function for cluster's dynamical evolution. 
But this physical process is proved to have insignificant influence on the mass-ratio distribution according to \citet{marks2011a} and also our test in \Cref{fig:p_q_lgeb}. 
Though in the functional expression of binding energy in \Cref{eq:11}, the mass ratio is one of the variables, we can see in the plane of binary period and mass ratio in \Cref{fig:p_q_lgeb}, where the colour depicts the value of binding energy, that change in mass-ratio hardly affects the value of binding energy. 
Thus, a process acting on the binding energy can barely modify the mass-ratio distribution.

Second, in the practice of our model, we find that the power-law indices of mass-ratio distribution in our model cannot be well-constrained by the data. 
The information of the mass-ratio distribution hides in the binary sequence for a simple stellar population (similar chemical composition). 
However, the stars in the solar neighbourhood have mixed metallicity, and this mixture wipes out the information of the mass-ratio distribution. 
It is more likely to constrain the distribution when we obtain the metallicity measurement for each star and split them into narrow metallicity bins \citep[e.g.,][]{liu2019}.
Therefore, we adopt the prior for $\gamma_1$, $\gamma_2$ and $q_{\mathrm{break}}$ from the wide binary results in \Cref{sec:app_q} to help constrain the mass-ratio distribution in our model.

\subsection{Metallicity-dependent selection bias and three-part IMF} \label{sec:dis_metal_bias}

An important consideration in our analysis is the potential bias introduced by our magnitude cut at $M_{\mathrm{G}} = 12.1$~mag. As noted by \citet{li2023}, \citet{kroupa2024}, etc., the stellar IMF may depend on metallicity. Since stellar luminosity at a given mass also varies with metallicity, a simple magnitude cut can lead to an underrepresentation of low-mass, high-metallicity stars in our sample. This effect can bias the inferred averaged IMF slopes if not properly accounted for.

\citet{kroupa2024} demonstrated that the IMF itself depends on metallicity according to:
\begin{equation}
\alpha_{1,2}=\alpha_{1,2,\mathrm{can}}+79.4(Z-Z_{\mathrm{MW}}),
\label{eq:kroupa_imf_z}
\end{equation}
where $\log_{10}(Z_{\mathrm{MW}}/Z_{\odot})=-0.10\pm0.05$ is the average metallicity of the solar neighbourhood stellar ensemble. Since we measure an average IMF over field stars with different metallicities, the combined effect of (1) metallicity-dependent stellar colors and (2) metallicity-dependent IMF slopes means that a simple horizontal magnitude cut will underestimate the number of low-mass, high-metallicity stars.

\begin{figure}
    \centering
    \includegraphics[width=0.75\linewidth]{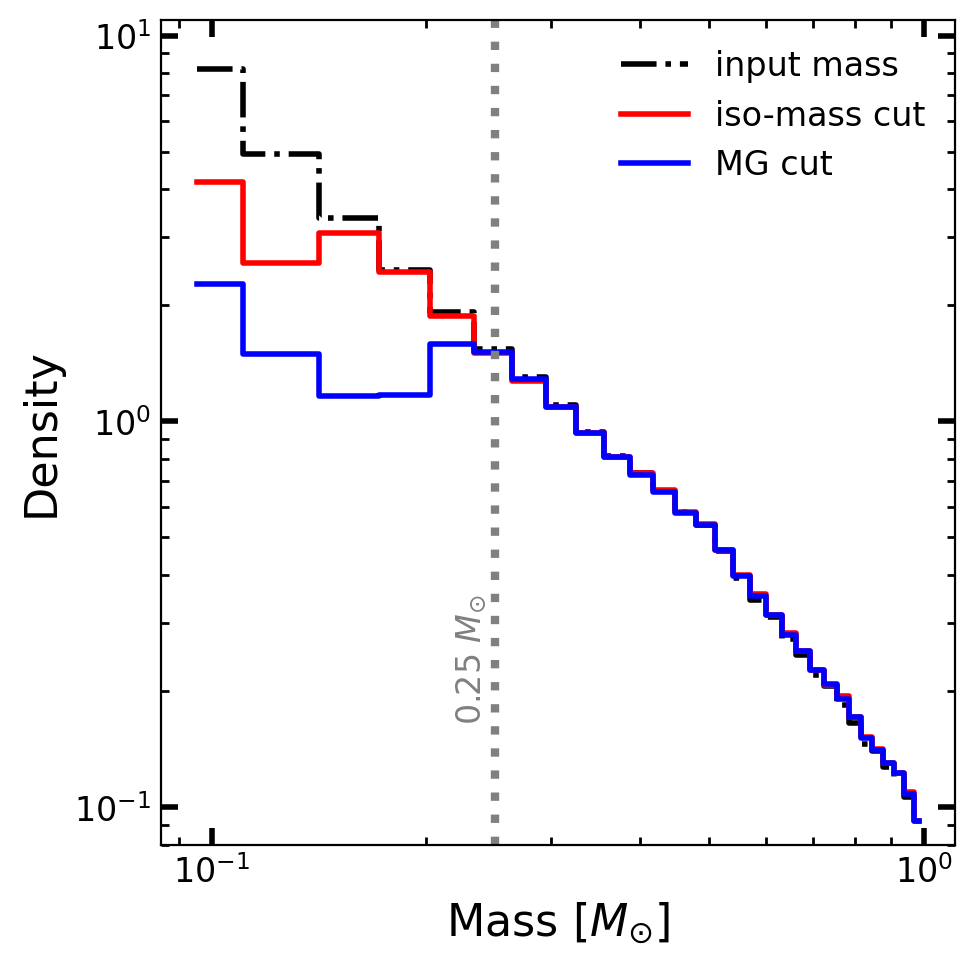}
    \caption{Mass distributions illustrating the effect of magnitude cuts when applying the metallicity-dependent IMF from \citet{kroupa2024}. The black dot-dashed line shows the input mass distribution before stellar evolution. The red solid line shows the mass distribution after our synthesis procedure and iso-mass cuts at $0.15~\mathrm{M_{\odot}}$ and $1.0~\mathrm{M_{\odot}}$. The blue solid line shows the mass distribution after the additional $M_{\mathrm{G}} = 12.1$~mag cut, revealing flattening below $0.25~\mathrm{M_{\odot}}$.
    The later two lines are shifted so that they overlap at $1~\mathrm{M_{\odot}}$ with the first line for easier comparison.}
    \label{fig:mass_dist_cut}
\end{figure}

To quantify this effect, we conducted mock data tests sampling metallicity from our derived Johnson's SU distribution (\Cref{sec:app_metal}) and masses according to the metallicity-dependent IMF of \citet{kroupa2024}. \Cref{fig:mass_dist_cut} shows the resulting mass distributions. After applying our magnitude cut at $M_{\mathrm{G}} = 12.1$~mag, the mass distribution becomes flattened below $0.25~\mathrm{M_{\odot}}$, while remaining nearly unchanged above this mass.

Since our forward modeling approach requires sampling across an extended mass range to properly synthesize binary systems, we cannot simply restrict our analysis to $m > 0.25~\mathrm{M_{\odot}}$. Instead, we adopt a three-part power-law IMF parameterization (\Cref{eq:6}), with $\alpha_0$ for the mass range $0.08$--$0.25~\mathrm{M_{\odot}}$, and our standard $\alpha_1$, $m_{\mathrm{break}}$, and $\alpha_2$ for higher masses. This formulation allows $\alpha_0$ to absorb the bias in the lowest mass bin, while $\alpha_1$ and $\alpha_2$ remain unbiased measures of the composite IMF in the solar neighbourhood for masses above $0.25~\mathrm{M_{\odot}}$.

We note that our derived IMF represents an average over the mixed stellar populations in the solar neighbourhood, each potentially having different metallicity-dependent IMFs as described by \citet{kroupa2024}. The values of $\alpha_1$ and $\alpha_2$ we report should therefore be interpreted as composite slopes reflecting this population mixture.

Finally, we acknowledge an important caveat regarding our imposed flux-limit at $M_{\mathrm{G}} = 12.1$~mag. 
An imposed flux-limit will over-represent binary systems in the sample, as unresolved binary systems appear brighter than single stars of the same mass, making them more likely to pass a magnitude cut. 
Consequently, our flux-limited sample may contain a higher fraction of (unresolved) binaries relative to single stars than would be present in a truly volume-complete sample. 
While our forward modeling explicitly accounts for the presence of unresolved binaries and aims to disentangle their effects, we caution that the flux-limit introduces a selection bias that preferentially includes binary systems. 
This could potentially affect our inferred binary fraction and IMF parameters. Similarly, parallax-limited samples might lead to analogous effects if binaries have poorer astrometric solutions and are thus excluded from the final sample, though we do not apply strict parallax cuts beyond the GCNS selection criteria. We find it difficult to avoid applying a flux-limit in our sample selection, as the PARSEC mass-luminosity relations do not reliably fit our sample for stars fainter than $M_{\mathrm{G}} = 12.1$~mag.

\subsection{Solutions in different distance bins} \label{sec:dis_distance}

% \kp{2. use the model on the sample 0-20 pc, 20-40 pc, 40-60 pc, 80-100 pc, to ensure the validity of the results; The difference between Kroupa's LFs}

In this section, we apply our model on the 0-20 pc, 20-40 pc, 40-60 pc, 60-80 pc and 80-100 pc subsamples, respectively. If the stellar population is sufficiently mixed in the 100-pc solar neighbourhood, our model should give invariable IMF parameters and binary fraction. This test will validate the robustness of our model and the uniformity of the stellar population in the solar neighbourhood.

\begin{figure*}
    \centering
    \includegraphics[width=0.8\linewidth]{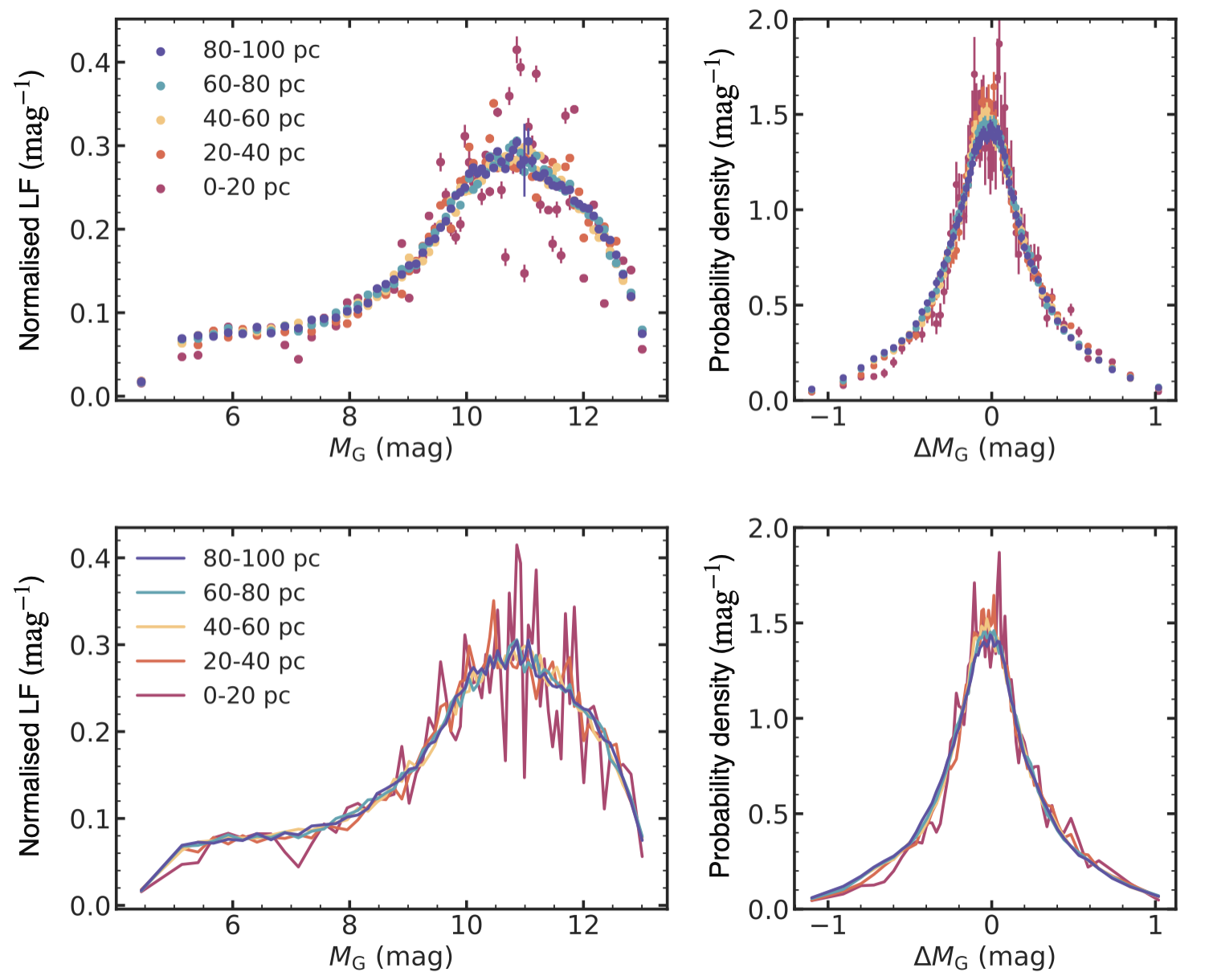}
    \caption{Luminosity functions and $\Delta M_{\mathrm{G}}$ distributions for GCNS sample in different distance bins. Top panel: drawn with dots and error bars for each bin. Bottom panel: drawn with lines which facilitates to observe the overall trend.}
    \label{fig:LF_var}
\end{figure*}

For the first step, we display the luminosity functions (LF) and $\Delta M_{\mathrm{G}}$ distributions for the different distance bins in \Cref{fig:LF_var}. We test for the consistency of distributions for each bin with the 80-100 pc LF and $\Delta M_{\mathrm{G}}$ distribution. By K-S test, we find that we cannot distinguish the LFs for each subsample with 95\% confidence, while each $\Delta M_{\mathrm{G}}$ distribution significantly deviates from the 80-100 pc $\Delta M_{\mathrm{G}}$ distribution. 

As distance goes farther, a larger proportion of binary systems become unresolved due to the limited angular resolution of \textsl{Gaia}, which theoretically should lead to a luminosity function with more bright sources and a more significant binary sequence ($\Delta M_{\mathrm{G}} \simeq -0.75$ mag) in the $\Delta M_{\mathrm{G}}$ distribution. This effect can be observed in \Cref{fig:LF_var}, where the binary sequence becomes more pronounced in the farther distance bins.

According to \citet{kroupa2001}, their nearby LF, $\phi_{\mathrm{near}}$ (from the 5.2-pc volume-limited trigonometric data), and their photometric LF, $\phi_{\mathrm{phot}}$ (from deep photometric pencil-beam observation), differ significantly for stars fainter than $M_{\mathrm{V}} \simeq 11.5~\mathrm{mag}$. \citet{kroupaUnificationNearbyPhotometric1995} explained that this discrepancy originates from the effect of unresolved multiple systems.
For our GCNS sample, unresolved binary do not affect the faint end of the luminosity function as strong as \citet{kroupaUnificationNearbyPhotometric1995} suggested, we suspect that the difference in LFs $\phi_{\mathrm{near}}$ and $\phi_{\mathrm{phot}}$ may be due to the combined influence of large statistical uncertainties induced by the small size of dataset in the derivation of $\phi_{\mathrm{near}}$, Malmquist bias caused by the using of photometric parallax in the derivation of $\phi_{\mathrm{phot}}$, and effect of unresolved binaries especially for the 5.2-pc volume-limited trigonometric data with its poorer angular resolution respect to \textsl{Gaia} DR3. 

The increasing proportion of unresolved binaries can be more easily detected in $\Delta M_{\mathrm{G}}$ distribution. It means that it is hard to constrain binary parameters sorely from luminosity functions, while the effect of unresolved binaries can be more accented in $\Delta M_{\mathrm{G}}$ space.

\begin{figure*}
    \centering
    \includegraphics[width=1\linewidth]{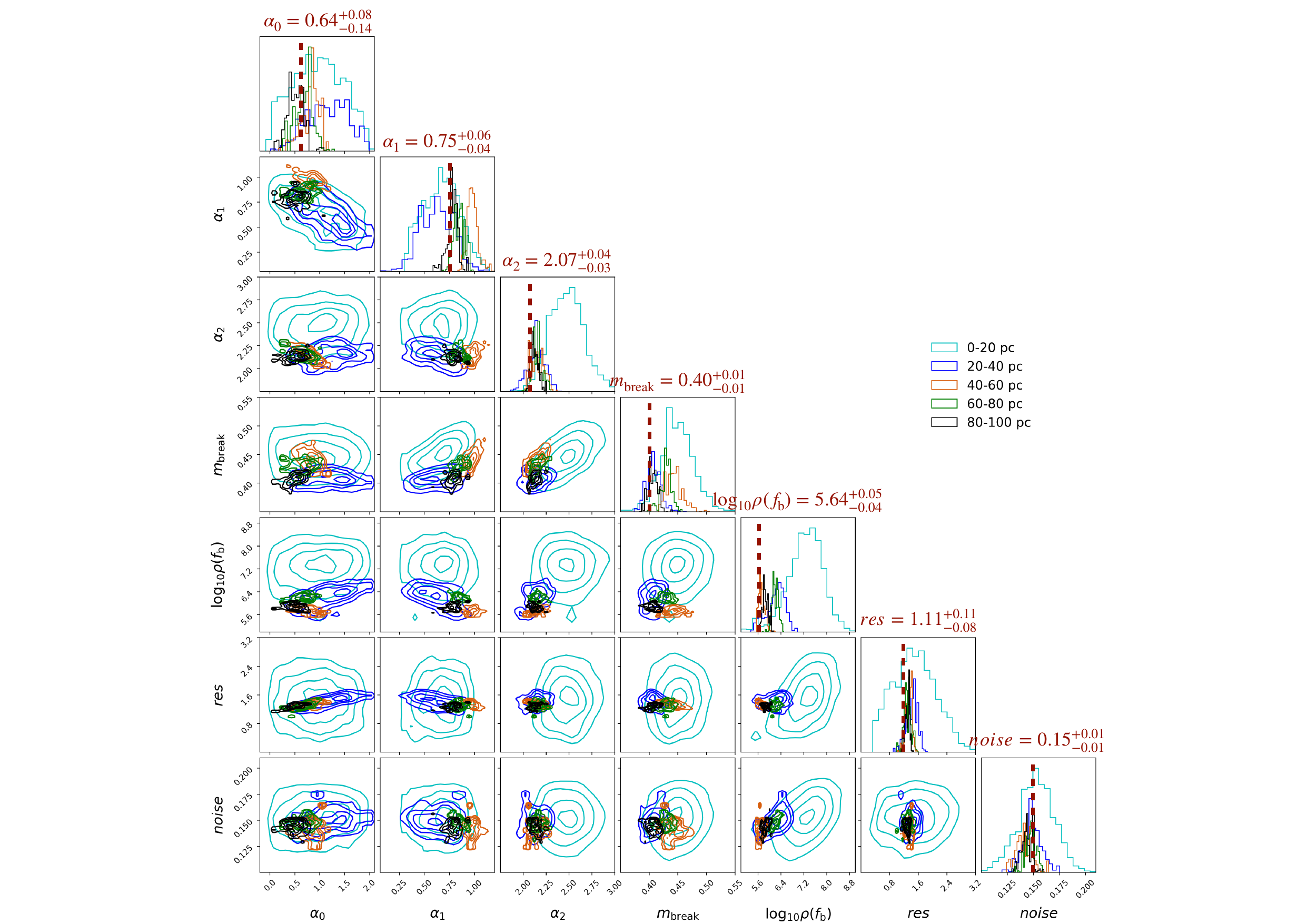}
    \caption{The derived parameter posteriors with our model in different distance bins. The brown vertical dashed lines and titles show the results from the whole 0-100 pc main sample which is demonstrated before in \Cref{sec:res} and \Cref{fig:mcmc}.}
    \label{fig:mcmc_d}
\end{figure*}

In \Cref{fig:mcmc_d}, we illustrate the derived parameters with our model in different distance bins. The corner plot in \Cref{fig:mcmc_d} shows 0.5, 1.0, 1.5, 2.0 $\sigma$ levels of each parameter derived in each distance bin. The uncertainties of each model parameter decline with increasing distance because of the larger volume of data for the farther distance bins. Among these distance bins, all parameters are indistinguishable within the 2-sigma uncertainty level, which shows the robustness of our population synthesis model, and indicates that that the stars in the solar neighbourhood are well-mixed. We observe that for stars within 0-20 pc, the parameters $\alpha_2$, $m_{\mathrm{break}}$, and $\log_{10}(\rho/\mathrm{M_{\odot}~pc^{-3}})$ are slightly higher than those in other distance bins. The larger values of $\alpha_2$ and $m_{\mathrm{break}}$ are likely due to the effect of saturation in photometric observations, causing \textsl{Gaia} to miss a portion of bright stars at such close distances \citep{gaiacollaborationGaiaDataRelease2022}.
This can also be illustrated by our completeness test in \Cref{fig:detec_rate}, where the completeness for bright stars at 0-20 pc is lower than that at farther distances. As a result, we tend to lose more massive stars, leading to a steeper IMF slope $\alpha_2$ and a higher break mass $m_{\mathrm{break}}$.
For the same reason, we tend to lose bright unresolved binary sources. Furthermore, given the higher binary fraction for brighter and more massive primary stars, it is reasonable to obtain a lower present-day binary fraction (i.e., higher $\log_{10}(\rho/\mathrm{M_{\odot}~pc^{-3}})$) for the nearby sample.

\subsection{Solution for the brighter end} \label{sec:dis_brightness}

% \kp{3. use the model on the brighter and fainter end respectively, primary stars with higher mass should have higher binarity}

It has been studied that the mean frequency of stellar companions increases from solar-type MS primaries to O-type MS primaries \citep{moe2017}. 
This observed trend in binary fraction with primary mass can also be explained by differences in age and dynamical processing apart from initial conditions \citep{kroupa2024}. 
Massive primaries are typically observed when young, retaining their primordial binaries, while low-mass field stars are older and have been dynamically processed such that many binaries have been disrupted. 

\begin{figure*}
    \centering
    \includegraphics[width=0.65\linewidth]{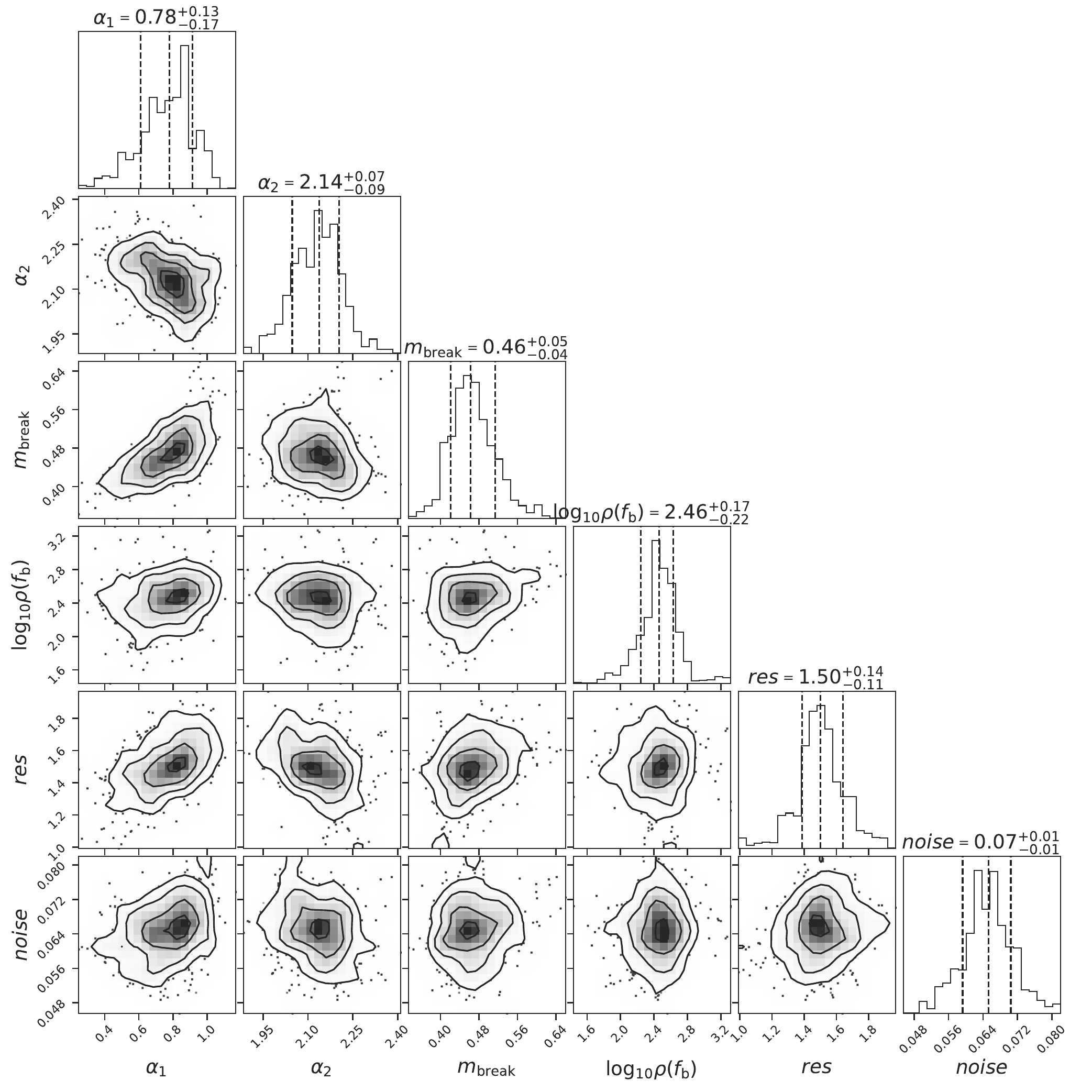}
    \caption{The posterior distribution of the model parameters for stars with $M_{\mathrm{G}}\leq 8$~mag. The best estimation of $\mathrm{log_{10}}(\rho/\mathrm{M_{\odot}~pc^{-3}})=2.46$ is equivalent to a present-day binary fraction of about $60\%$.}
    \label{fig:mcmc_bright}
\end{figure*}

\begin{figure*}
    \centering
    \includegraphics[width=0.8\linewidth]{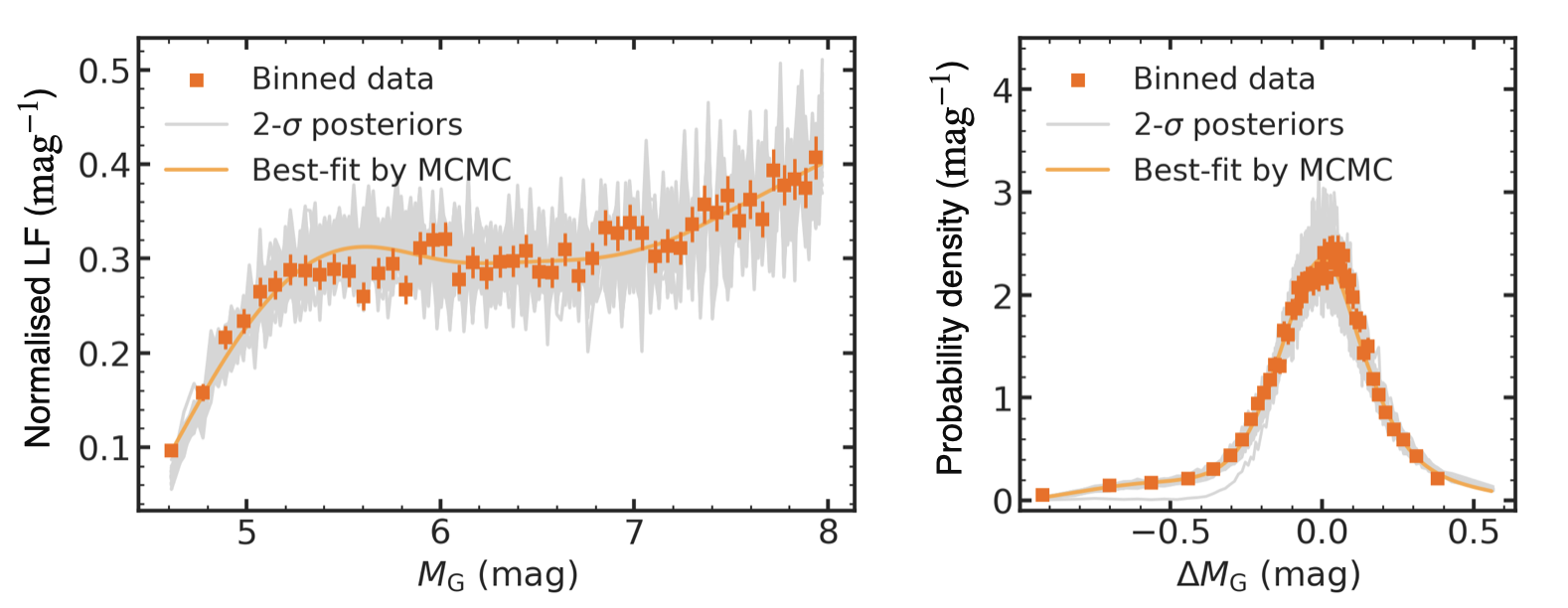}
    \caption{Comparison between observed luminosity function $\phi^{\mathrm{bright}}_{\mathrm{obs}}(M_{\mathrm{G}})$, $\Delta M^{\mathrm{bright}}_{\mathrm{G,\mathrm{obs}}}$ distribution and our best-fit model.}
    \label{fig:fitting_bright}
\end{figure*}

We could expect, if we separate our samples by luminosity into a bright part and a faint part, the present-day binary fraction of the brighter part should be higher than the fainter sub-sample. We define the brighter part as stars with $M_{\mathrm{G}}\leq 8$~mag, and the posteriors of the model parameters and the fitting result is displayed in \Cref{fig:mcmc_bright} and \Cref{fig:fitting_bright}. Since we only use the brighter part of the sample, there's no need to adapt the three-part IMF to avoid the bias at the low-mass end. Therefore, we only fit a two-part IMF for this sub-sample, and only $\alpha_2$ is constrained in this case.
The resultant IMF parameters is consistent with the result derived with the whole sample within 2-$\sigma$ level, while the binary fraction is derived to be higher, which is approximately $60\%$, compared to our main result. The derived $noise$, which corresponds to the standard deviation of the Gaussian noise added to the simulated $\Delta M_{\mathrm{G}}$ distribution, is smaller than our main result. This is reasonable, as the observational uncertainty should be relatively small for brighter stars. This result for the stars at the brighter end once again demonstrates the reliability of our population synthesis model.

\subsection{The $\xi$ parameter to compare different IMF measurements} \label{sec:xi}

\citet{martin-navarro2019} introduces a new quantity, $\xi$, for an unbiased comparison among
different IMF parametrizations. The $\xi$ parameter quantifies the mass fraction locked in low-mass stars, and is defined as

\begin{equation}
\xi \equiv \dfrac{\int_{0.2}^{0.5}{m \cdot X(m)dm}}{\int_{0.2}^{1.0}{m \cdot X(m)dm}},
\label{eq:mar}   
\end{equation}
where $X(m)$ is the IMF expressed in linear mass units.

\citet{martin-navarro2019} has calculated that for a \citet{chabrierGalacticDiskMass2003} IMF, $\xi=0.4607$; and for \citet{salpeter1955} IMF, $\xi=0.6370$. \citet{kroupa1993} has provided the parameter uncertainty for his IMF, so we can obtain the corresponding $\xi=0.5194_{-0.0585}^{+0.0581}$. For the IMF derived in this paper, $0.5070_{-0.0096}^{+0.0068}$, with much smaller uncertainty. All $\xi$ values are summarized in \Cref{tab:xi}.

\begingroup
\setlength{\tabcolsep}{10pt} % Default value: 6pt
\renewcommand{\arraystretch}{1.25} % Default value: 1
\begin{table}
  \caption{The $\xi$ parameter of different IMF measurements.}
  \label{tab:xi}
  \begin{tabular}{ccc}
    \hline
    $\xi$ & Reference \\
    \hline
    $0.6370$ & \citet{salpeter1955} \\
    $0.5194_{-0.0585}^{+0.0581}$ & \citet{kroupa1993} \\
    $0.4607$ & \citet{chabrierGalacticDiskMass2003} \\
    $0.5070_{-0.0096}^{+0.0068}$ & This work \\
    \hline
  \end{tabular}
\end{table}
\endgroup

\subsection{Caveats on mass-luminosity relations from PARSEC isochrones} \label{sec:dis_parsec}

We acknowledge an important caveat regarding the mass-luminosity relations (MLRs) from PARSEC isochrones used in our analysis. As highlighted by \citet{kroupa2024}, nowadays stellar isochrones may yield inaccurate MLRs for low-mass stars, which could introduce biases in the inferred IMF slopes $\alpha_1$ and $\alpha_2$. Our choice of PARSEC was driven by its metallicity-dependent MLRs, which are essential for modeling our stellar population with metallicity variations spanning $-1.0 < \mathrm{[M/H]} < 0.6$~dex.

\begin{figure}
    \centering
    \includegraphics[width=0.85\linewidth]{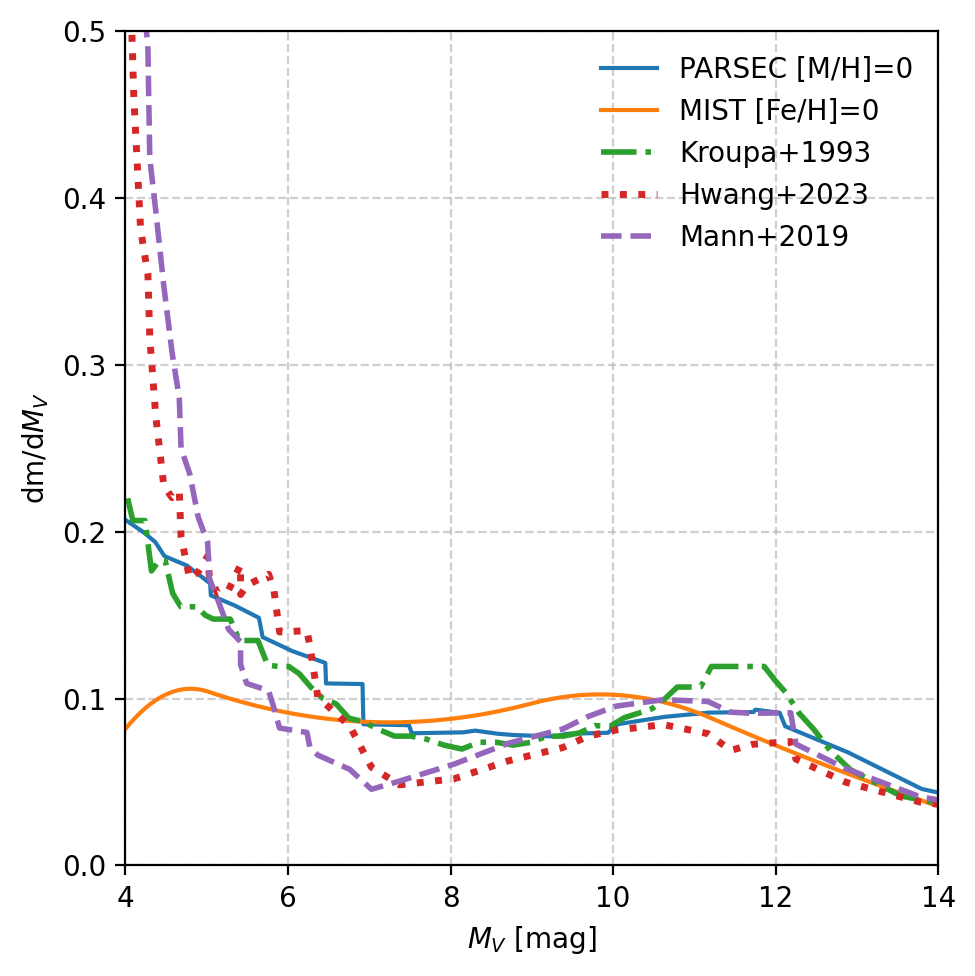}
    \caption{Comparison of the derivatives of mass-luminosity relations ($\mathrm{d}m/\mathrm{d}M_V$) from different stellar evolution models: PARSEC 1.2s ([M/H]=0), MIST ([Fe/H]=0), \citet{kroupa1993}, \citet{hwang2023}, and \citet{mann2019}. 
    The differences among models highlight the systematic uncertainties in MLRs for low-mass stars, which is also shown in \citet{kroupatout1997}.
    The relations are converted to $V$-band for consistency.}
    \label{fig:dmdl_models}
\end{figure}

\begin{figure}
    \centering
    \includegraphics[width=0.85\linewidth]{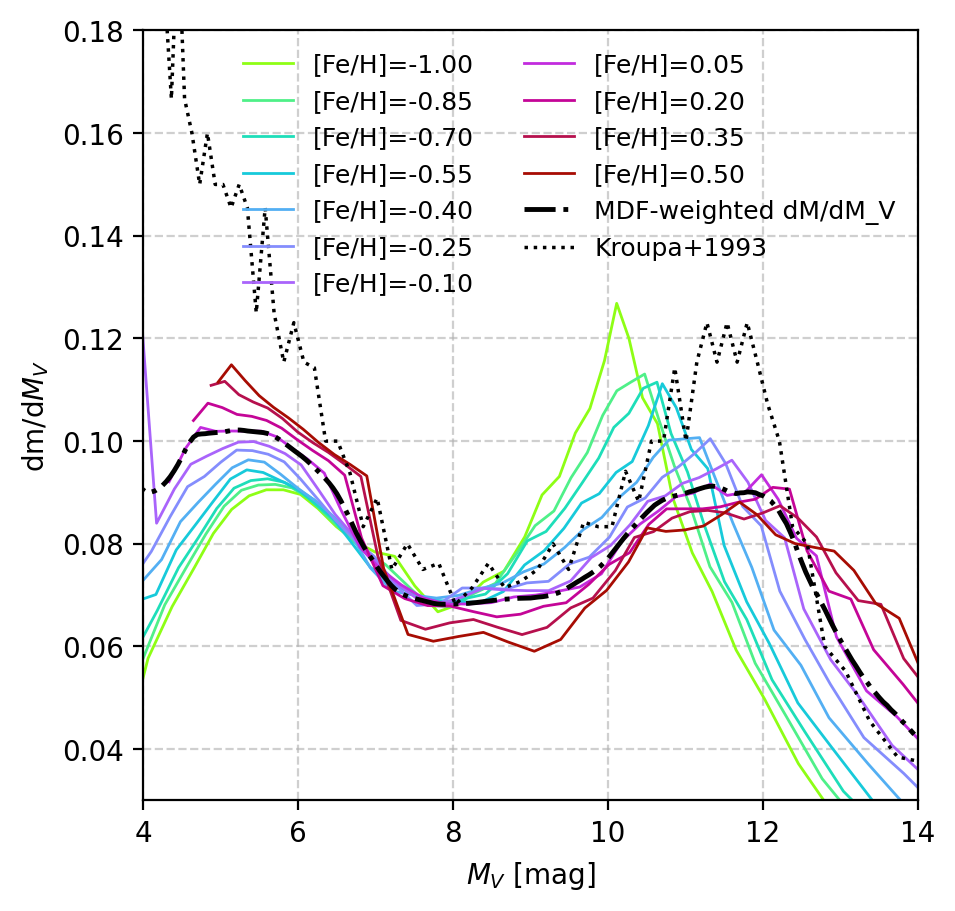}
    \caption{Derivatives of mass-luminosity relations from PARSEC 1.2s at different metallicities (colored lines), the metallicity-weighted average using our derived solar neighbourhood metallicity distribution (thick line), and the \citet{kroupa1993} relation (black line) for reference. 
    PARSEC's metallicity-dependent MLRs are crucial for our analysis of the mixed stellar population.
    The relations are converted to the $V$ band for consistency.}
    \label{fig:dmdl_parsec}
\end{figure}

While we recognize PARSEC's limitations \citep[e.g., calibration studies using open clusters on PARSEC isochrones][]{wangfan2025}, several factors support its use in our context. 
First, \citet{hwang2023} demonstrated that PARSEC better reproduces the main-sequence ridge in mass-CMD relations derived from wide-binary dynamical masses compared to MIST, the primary alternative stellar evolution model with metallicity dependence. 
Second, our analysis confirms that PARSEC's $\mathrm{d}m/\mathrm{d}M_{\mathrm{G}}$ distribution matches the observed luminosity function profile in the range $M_{\mathrm{G}} = 4.1$--$12.1$~mag, particularly reproducing the two characteristic broad bumps at $M_{\mathrm{G}} \approx 5.5$ and $10.75$~mag (see \Cref{fig:dmdl_models,fig:dmdl_parsec}).

\Cref{fig:dmdl_models} compares the derivatives of mass-luminosity relations from different models: PARSEC 1.2s ([M/H]=0), MIST ([Fe/H]=0), \citet{kroupa1993}, \citet{hwang2023}, and \citet{mann2019}. \Cref{fig:dmdl_parsec} shows PARSEC's $\mathrm{d}m/\mathrm{d}M_V$ (Here $M_V$ denotes the absolute magnitude in the Johnson $V$ band, and is converted from \textit{Gaia}'s $M_{\mathrm{G}}$) for different metallicities, along with the metallicity-weighted average (using our derived metallicity distribution) and the \citet{kroupa1993} relation for comparison.

Nevertheless, we recognize that recalculating with alternative MLRs would be computationally prohibitive, and many available MLRs lack the necessary metallicity dependence. Future work with improved stellar models—particularly those that have been empirically calibrated for low-mass stars across a range of metallicities—will be valuable for refining IMF determinations. We caution that systematic uncertainties in PARSEC's MLRs may propagate into our inferred IMF slopes, though the consistency of our results with canonical values suggests these effects are not dominant within our measurement precision.

\section{Conclusions} \label{sec:conclusion}

In this paper, we remeasure the stellar initial mass function for field stars in the 100-pc solar neighbourhood in the mass range $0.25<m/\mathrm{M_{\odot}}<1.0$. With the most up-to-date volume-complete trigonometric parallax data from \textsl{Gaia} DR3, this measurement establishes a new IMF benchmark for field stars in the solar neighbourhood. In our forward model for the IMF, we redefine the IMF as the mass distribution according to which stars form with 100\% binarity and are randomly paired at birth, then undergo dynamical evolution similar to that in a star cluster. The consequent binary fraction and mass-ratio distribution after evolution are compared with the observed present-day binary population properties to constrain our model parameters, including $\alpha_0$, $\alpha_1$, $\alpha_2$, $m_\mathrm{break}$ (three-part power-law IMF), $\gamma_1$, $\gamma_2$, $q_{\mathrm{break}}$ (mass-ratio distribution), $res$ (angular resolution of \textsl{Gaia}), and $\mathrm{log_{10}}(\rho/\mathrm{M_{\odot}~pc^{-3}})$ (typical cluster density for the cluster-mode dynamical evolution, related to present-day binary fraction).

Due to the population mixture (in metallicity and binary properties) in the data, the present-day mass-ratio distribution cannot be well constrained by the GCNS dataset alone. Therefore, we fix the parameters of the mass-ratio distribution using information provided by wide binaries compiled by \citet{el-badryMillionBinariesGaia2021}. We characterize the present-day mass-ratio distribution as a combination of a piece-wise power-law function and an excess of twin binaries, obtaining $\gamma_1=1.89_{-0.20}^{+0.23}$, $\gamma_2=0.20_{-0.13}^{+0.12}$, $q_{\mathrm{break}}=0.44_{-0.02}^{+0.02}$, and $F_{\mathrm{twin}}=0.08_{-0.01}^{+0.01}$. We apply these as Gaussian priors in our Bayesian inference.

% \kp{2. Main results: IMF (consistent with previous studies, but smaller uncertainties), binary fraction and Gaia resolution}

Our best estimate for the averaged field star IMF is $\alpha_1=0.75^{+0.06}_{-0.04}$, $\alpha_2=2.07^{+0.04}_{-0.03}$ and $m_{\mathrm{break}}=0.40^{+0.01}_{-0.01}$ for stars with $0.25<m<1.0~M_{\odot}$, when applying the mass-ratio distribution priors above, which is consistent with Kroupa's canonical IMFs (from the $\xi$-parameter perspective) but with substantially smaller uncertainties, owing to the unprecedented coverage and quality of the \textsl{Gaia} astrometric data. A caveat is that IMF measurements are subject to not yet fully understood systematics due to reliance on theoretical stellar mass-luminosity relations.
Additionally, our model provides a measurement of the binary fraction in the solar neighbourhood. The density of the characteristic cluster in our model is $\mathrm{log_{10}}(\rho/\mathrm{M_{\odot}~pc^{-3}})=5.64^{+0.05}_{-0.04}$, which corresponds to a fitted present-day binary fraction of approximately 26\%. Note that we consider only binary systems to obtain this value. Furthermore, we obtain the average angular resolution of \textsl{Gaia} DR3 to be $res=1.11^{+0.11}_{-0.08}$ arcsec, representing Gaia DR3's capability to resolve objects separated by this angular distance.

% \kp{3. Main results from discussion}

In the discussion, we first examine the luminosity functions and $\Delta M_{\mathrm{G}}$ distributions for stars in different distance bins: 0-20 pc, 20-40 pc, 40-60 pc, 60-80 pc, and 80-100 pc. We perform K-S tests to check whether these luminosity functions and $\Delta M_{\mathrm{G}}$ distributions are consistent with each other. The K-S tests confirm that luminosity functions for different distance bins cannot be distinguished from one another, whereas $\Delta M_{\mathrm{G}}$ distributions for different distance bins are drawn from different distributions. The results of our K-S tests are understandable because: (1) The form of the luminosity function depends on the IMF and mass-luminosity relation. If the stellar population in the 100-pc solar neighbourhood is well-mixed, we should obtain consistent IMFs for different distance bins. (2) The profile of the $\Delta M_{\mathrm{G}}$ distribution depends primarily on the properties of the binary population, i.e., unresolved binary fraction and present-day mass-ratio distribution. Since it becomes more difficult for \textsl{Gaia} to resolve binaries with increasing distance, the proportion of unresolved binaries increases, causing variation in the $\Delta M_{\mathrm{G}}$ distribution. With the angular resolution of \textsl{Gaia} taken into account in our model, we expect to recover similar overall binary fractions for different distance bins.
Our model is thus validated by yielding compatible IMF and binary fraction parameters among different distance bins.

We further test our model by applying it only to the brighter end of the CMD for stellar masses $0.5<m/\mathrm{M_{\odot}}<1.0$. Our model's robustness is further confirmed by the compatible IMF parameters and higher binary fractions revealed for systems with more massive primary stars.

Finally, we provide the $\xi$ parameter for our new IMF to facilitate comparison with measurements of other stellar populations, which is $\xi=0.5070_{-0.0096}^{+0.0068}$.

\section*{Acknowledgements}
% Entry for the table of contents, for this guide only
\addcontentsline{toc}{section}{Acknowledgements}

The authors appreciate the valuable comments and suggestions from the reviewer.

The authors appreciate discussions with Zhiqiang Yan, Dan Qiu, Hans-Walter Rix, and Long Wang. 

This work is supported by the National Key R\&D Program of China No. 2025YFF0511000.

This work used the data from the European Space Agency (ESA) mission Gaia (\url{https://www.cosmos.esa.int/gaia}), processed by the Gaia Data Processing and Analysis Consortium (DPAC; \url{https://www.cosmos.esa.int/web/gaia/dpac/consortium}). Funding for the DPAC has been provided by national institutions, in particular the institutions participating in the Gaia Multilateral Agreement. Guoshoujing Telescope (the Large Sky Area Multi-Object Fiber Spectroscopic Telescope LAMOST) is a National Major Scientific Project built by the Chinese Academy of Sciences. Funding for the project has been provided by the National Development and Reform Commission. LAMOST is operated and managed by the National Astronomical Observatories, Chinese Academy of Sciences. 

\textsl{Facilities}: \textsl{Gaia}, LAMOST.

\textsl{Software}: \texttt{IPython} \citep{PER-GRA:2007}, \texttt{jupyter} \citep{2016ppap.book...87K}, \texttt{pandas} \citep{reback2020pandas}, \texttt{Astropy} \citep{2022ApJ...935..167A}, \texttt{numpy} \citep{harris2020array}, \texttt{scipy} \citep{2020SciPy-NMeth}, \texttt{matplotlib} \citep{Hunter:2007}, \texttt{PyTorch} \citep{paszke2017automatic}, \texttt{emcee} \citep{2013ascl.soft03002F}.

%%%%%%%%%%%%%%%%%%%%%%%%%%%%%%%%%%%%%%%%%%%%%%%%%%

\section*{Data Availability}

The data underlying this article are available online. The datasets were derived from sources in the public domain: The Gaia Data Archive \url{https://gea.esac.esa.int/archive/}, and National Astronomical Data Center \url{https://nadc.china-vo.org}.



\bibliographystyle{mnras}
\bibliography{IMF} % if your bibtex file is called example.bib

%\begin{thebibliography}{99}
%\end{thebibliography}

%%%%%%%%%%%%%%%%%%%%%%%%%%%%%%%%%%%%%%%%%%%%%%%%%%

%%%%%%%%%%%%%%%%% APPENDICES %%%%%%%%%%%%%%%%%%%%%

\appendix

\section{Completeness analysis of GCNS} \label{sec:app_vmax}

In this appendix, we describe our approach to assessing the completeness and detection rate of the GCNS catalog. Since the selection function shows minimal variation with colour $G_{\mathrm{BP}}-G_{\mathrm{RP}}$ \citep{rix2021}, we focus our analysis on the absolute magnitude $M_{\mathrm{G}}$ and distance dependencies.

\subsection{Detection rate methodology}

Following \citet{rybizki2018} and \citet{mateu2020}, we quantify the detection rate (completeness) by comparing the observed stellar distribution to control samples generated with different assumed density profiles. This approach allows us to assess both the completeness of our sample and the model dependence of our conclusions.

We generate two types of control samples: (1) constant density and (2) exponential disk density. The exponential disk density distribution is given by:
\begin{equation}
\rho(R_{\mathrm{xy}},~z) = \rho_0 \exp\left(-\frac{R_{\mathrm{xy}}}{h_R}\right) \exp\left(-\frac{|z|}{h_z}\right),
\label{eq:exp_disk_app}
\end{equation}
where $\rho_0$ is the central density, $R_{\mathrm{xy}}$ is the Galactocentric distance along the Galactic plane, $z$ is the height above the Galactic plane, $h_{\mathrm{R}} = 3$~kpc is the scale length, and $h_{\mathrm{z}} = 250$~pc is the scale height \citep{danmaoz2017,tang2024}. The spatial position of the sun is set to $(X_{\odot},~Y_{\odot},~Z_{\odot})=(-8.35,~0,~0.01)$~kpc \citep{Griv2021}.

The GCNS catalogue is divided into sliding bins in $M_{\mathrm{G}}$ with bin center steps of 0.5 mag and bin width of 1 mag. Since observed stars obey a luminosity function, different $M_{\mathrm{G}}$ bins contain different intrinsic numbers of stars. To make detection rates comparable across bins, we calculate a normalization factor:
\begin{equation}
C(M_{\mathrm{G}})=\dfrac{N^{\Delta d}_{\mathrm{obs}}(M_{\mathrm{G}})}{N^{\Delta d}_{\mathrm{ctrl}}},
\label{eq:norm_factor_app}
\end{equation}
where $N^{\Delta d}_{\mathrm{obs}}(M_{\mathrm{G}})$ is the number of observed stars in a given $M_{\mathrm{G}}$ bin within the reference distance range $\Delta d = 30$--60~pc, and $N^{\Delta d}_{\mathrm{ctrl}}$ is the corresponding number in the control sample. We choose this reference range because it is expected to be complete for our sample.

Within each $M_{\mathrm{G}}$ bin, we further divide the sample into distance bins with width $\Delta d_{\mathrm{bin}} = 10$~pc. The detection rate for each $(M_{\mathrm{G}}, d)$ bin is then:
\begin{equation}
P(M_{\mathrm{G}},d)=\dfrac{1}{C(M_{\mathrm{G}})}\dfrac{N_{\mathrm{obs}}(M_{\mathrm{G}},d)}{N_{\mathrm{ctrl}}(d)},
\label{eq:detect_rate_app}
\end{equation}
where $N_{\mathrm{obs}}(M_{\mathrm{G}},d)$ is the observed number of stars and $N_{\mathrm{ctrl}}(d)$ is the expected number from the control sample in the given bin.

\subsection{Results and discussion}

\Cref{fig:detec_rate} in the main text shows the detection rates as a function of $M_{\mathrm{G}}$ and distance for both density models. Several key features emerge from this analysis:

\textit{Magnitude dependence}: For fainter stars, there is no systematic trend in detection rate along $M_{\mathrm{G}}$ axis for neither density model. This indicates that our sample does not suffer from significant absolute magnitude-dependent incompleteness in the $M_{\mathrm{G}}$ range we focus on: $4.1 < M_{\mathrm{G}} < 12.1~\mathrm{mag}$.

\textit{Distance dependence}: A trend in detection rate with distance appears when assuming the constant density model but not for the exponential density model. This indicates that the effect of exponential disk structure becomes noticeable when distance goes to 60~pc as shown in \Cref{fig:detec_rate}. However, as this distance dependence is orthogonal to the $M_{\mathrm{G}}$ axis, it will not bias our IMF determination.

\textit{Nearby bright stars}: The detection rate for nearby bright stars (approximately $d < 30$~pc and $M_{\mathrm{G}} < 6$~mag, corresponding to the lower-left region of \Cref{fig:detec_rate}) is slightly lower than elsewhere. This deficiency is likely caused by flux saturation in \textit{Gaia}'s photometry for bright stars \citep{collaborationGaiaEarlyData2021}, and the small number statistics of Poisson fluctuations providing us with a mode smaller than the truth. The affected region accounts for only approximately 0.6\% of the stars in our main analysis sample ($4.1 < M_{\mathrm{G}} < 12.1$~mag, $0 < d < 100$~pc).

Since only trends along the $M_{\mathrm{G}}$ axis would significantly bias our IMF determination, and we find no such systematic trends, we conclude that our sample is effectively complete for IMF analysis in the range $4.1 < M_{\mathrm{G}} < 12.1$~mag within 100~pc. To minimize model assumptions in our IMF derivation, we do not apply selection function corrections. This approach is conservative and avoids introducing additional systematic uncertainties from model-dependent correction factors.

\subsection{Effective volume formalism}

If the detection rate dose show systematic trends along the $M_{\mathrm{G}}$ axis, we can correct for incompleteness subsequently using the effective volume formalism. Following \citet{rix2021}, we define the effective volume $V_{\mathrm{eff}}$ for a given $M_{\mathrm{G}}$ as

\begin{equation}
V_{\mathrm{eff}}=\Omega \iint{\mathrm{d}z \mathrm{d}R \cdot R \rho(R,z) \cdot S(R,z)},
\label{eq:veff}
\end{equation}
where $\Omega$ is the solid angle covered by the data, $\rho(R,z)$ is the stellar density profile, and $S(R,z)$ is the selection function (i.e., detection rate) at position $(R,z)$.

The unbiased estimator for $\phi(M_{\mathrm{G}})$ will then be \Cref{eq:psi0} as described in \citet{tinneyFaintestStarsSchmidt1993}.

\begin{equation}
\phi(M_{\mathrm{G}})=\sum{\dfrac{1}{V_{\mathrm{a}}(M_{\mathrm{G}})}}
\label{eq:psi0}   
\end{equation}

\section{Metallicity distribution in the solar neighbourhood} \label{sec:app_metal}

~\\

% \kp{fit for metallicity sampling function, correct for LAMOST metallicity, correct for the selection effect on metallicity, extract intrinsic MDFs}

% \kp{data}

All stars in the solar neighbourhood ought to have very comparable metallicity distributions, except for the youngest stars that only formed recently, since they all share the same chemical enrichment history of the Galaxy. The metallicity distribution reflects the integrated history of star formation and chemical evolution in this region. This justifies our approach of applying a single averaged metallicity distribution function to our population synthesis model, as the field stars within 100~pc represent a well-mixed population with a common chemical heritage.

In this section, we discuss how we extract the intrinsic metallicity distribution (MDF; i.e. measurement uncertainty deconvolved MDF) in the solar neighbourhood that will be treated as a sampling function for the metallicity of our simulated stars. 

We make use of the metallicity measurement derived from the Data Release Nine (DR9) of LAMOST Low-Resolution Spectroscopic Survey (LRS) \footnote{\url{http://www.lamost.org/dr9/v2.0/}}. For A-, F-, G-, and K-type stars, we download the LAMOST LRS Stellar Parameter Catalog of A, F, G, and K Stars from the official website and use the metallicity provided in the catalog. \citet{niu2023} investigated the systematic errors of stellar chemical abundances in the LAMOST and Gaia databases. We use the correction function in Niu's paper to calibrate the LAMOST DR9 metallicity for A, F, G, K stars. 

The metallicity measurement for M-type stars is a more complicated problem, for their faint nature and our relatively poor understanding of their atmospheric model \citep{qiu2024}. \citet{qiu2024} empirically calibrate the LAMOST DR9 metallicity of M-type stars using wide binaries with a F-, G-, or K-dwarf primary and a M-dwarf companion, taking the systematic calibration in \citet{niu2023} into consideration. Since what we desire is an averaged MDF covering stars from $M_{\mathrm{G}}=4.1$ mag to $M_{\mathrm{G}}=12.1$ mag, we combine the LAMOST official AFGK-star catalog with metallicity corrected ourselves and the M-type star catalog provided by \citet{qiu2024} to perform the calculation.

\begin{figure*}
    \centering
    \includegraphics[width=0.8\linewidth]{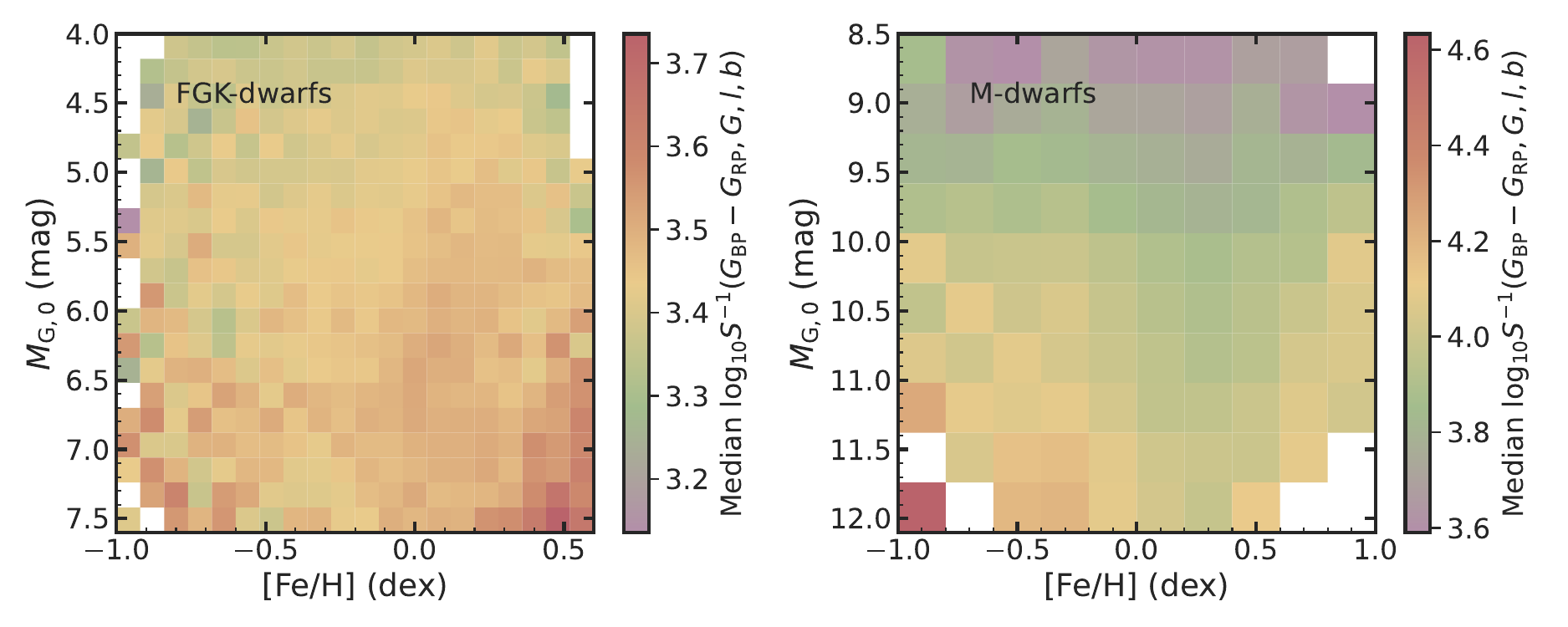}
    \caption{The distribution of inverted selection functions $\mathrm{log_{10}}S^{-1}(G_{\mathrm{BP}}-G_{\mathrm{RP}},G,l,b)$ of LAMOST DR9 AFGK-type and M-type catalogs.}
    \label{fig:lamost_sf}
\end{figure*}

In order to eliminate selection effects that might act on the two catalogs, we refer to the method in \citet{liuchao2017}, which recovered the selection function by comparing the distribution of the spectroscopic stars (here referred to as our LAMOST catalogs) with that of the photometric dataset (here referred to as \textsl{Gaia} DR3) in a 4-dimensional space of colour, apparent magnitude, and 2-D position. The selection function $S(c,m,l,b)$ is defined as

\begin{equation}
S(c,m,l,b)=\dfrac{n_{\mathrm{sp}}(c,m,l,b)}{n_{\mathrm{ph}}(c,m,l,b)},
\label{eq:sf}   
\end{equation}
where $c,m,l,b$ are respectively colour, apparent magnitude, Galactic longitude, and Galactic latitude. $n_{\mathrm{sp}}(c,m,l,b)$ is the stellar density of the spectroscopic survey, and $n_{\mathrm{ph}}(c,m,l,b)$ the stellar density of the spectroscopic survey at certain point of this 4-dimensional space. In order to improve the precision of the computation, \citet{liuchao2017} calculate $S^{-1}(c,m,l,b)$ instead of $S$ in their paper. Before we apply their procedure, we acquire the interstellar extinction from \citet{edenhofer2024} and compute the intrinsic absolute magnitude $M_{\mathrm{G,0}}$ with the geographical distance \texttt{rgeo} estimated by \citet{bailerjones2018} for stars in both catalogs. Then we cut our FGK-dwarf samples with $4.0<M_{\mathrm{G,0}}<7.6~\mathrm{mag}$ and $100<\texttt{rgeo}<300~\mathrm{pc}$, and cut the M-dwarf catalog with $8.5<M_{\mathrm{G,0}}<12.1~\mathrm{mag}$ and $20<\texttt{rgeo}<160~\mathrm{pc}$, to guarantee both the coverage of our interested stellar mass range and sample completeness. Then we perform Liu's program in equal-frequency bins of metallicity to avoid the bias caused by uneven relative Poisson error \citep{maizapellaniz2005a}. \Cref{fig:lamost_sf} demonstrates how the value $S^{-1}(c,m,l,b)$ varies with metallicity and absolute magnitude for FGK-dwarfs and M-dwarfs, respectively. We treat $S^{-1}(c,m,l,b)$ as counting weights of each star to correct for the selection effect of our LAMOST DR9 catalogs. 

We resample our data with $S^{-1}(c,m,l,b)$ (normalised to 0-1) as the possibility. In this way, stars suffered from severer selection effect will be more likely to be retained. Then, since the metallicity of FGK-dwarfs possess smaller measurement uncertainty compared to that of the M-dwarfs, we should first extract the intrinsic metallicity distributions of both kinds of samples. We choose the Johnson's SU distribution to fit for the intrinsic metallicity distributions. The Johnson's SU-distribution is a four-parameter probability distribution first investigated by N. L. Johnson in 1949 \citep{johnson1949}. It is a transformation of the normal distribution which can be easily sampled from: 

\begin{equation}
Y=\gamma + \delta \mathrm{sinh^{-1}} \left( \dfrac{X-\mathrm{loc}}{\mathrm{scale}} \right)
\label{eq:johnson}   
\end{equation}
where $Y\sim \mathcal{N}(0,1)$, and $X$ is the random variable that is Johnson's SU distributed. Considering this Johnson's SU intrinsic form coupled with an observational model perturbed by noise, the likelihood for each data point can be formalized as: 

\begin{align}
p(z_{\mathrm{obs},i}|\sigma^2_{z,i},\gamma,\delta,\mathrm{loc},\mathrm{scale})
\notag
\\=\int_{z_{\mathrm{min}}}^{z_{\mathrm{max}}}{\mathcal{N}(z_{\mathrm{obs},i}|z,\sigma^2_{z,i}) p_{\mathrm{intrinsic}}(z|\gamma, \delta, \mathrm{loc}, \mathrm{scale})}dz.
\label{eq:intrinsic_baye}   
\end{align}

The joint likelihood should be:

\begin{align}
\mathcal{L}=\prod_{i} p(z_{\mathrm{obs},i}|\sigma^2_{z,i},\gamma,\delta,\mathrm{loc},\mathrm{scale}),
\label{eq:intrinsic_baye_joint}   
\end{align}
where we use $z$ to stand for the intrinsic metallicity of each star, and $z_{\mathrm{obs}}$ as the observational metallicity in our catalogs which is perturbed by observational uncertainty $\sigma_{z,i}$.

\begin{figure*}
    \centering
    \includegraphics[width=0.7\linewidth]{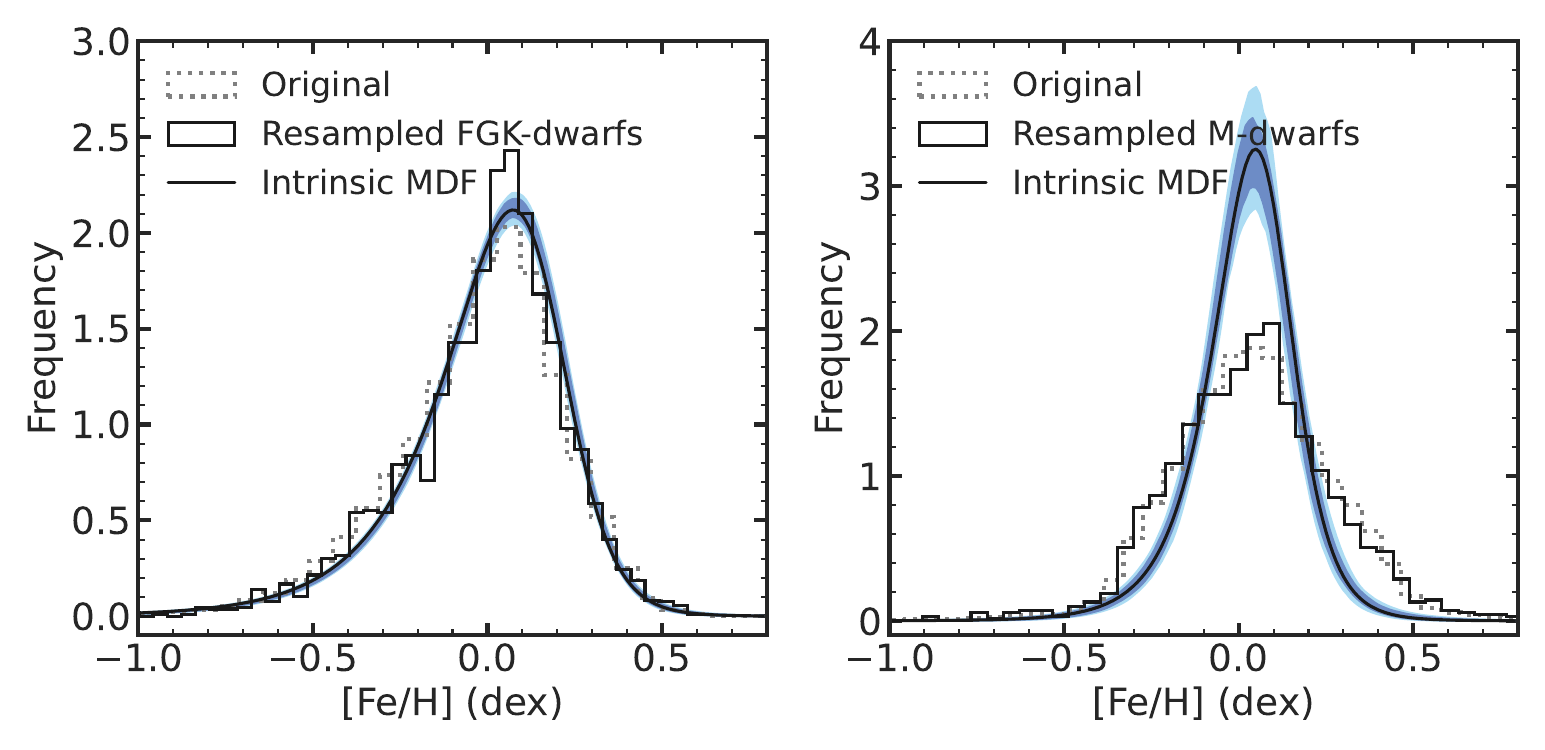}
    \caption{Original and resampled FGK- and M-dwarf's metallicity distributions. one- and two-sigma posteriors of the intrinsic metallicity distributions are drawn with deep blue and light blue shaded areas. The best fit Johnson's SU functions of the  intrinsic metallicity distributions are plotted with solid black lines in both plots.}
    \label{fig:fgkm_mdfs}
\end{figure*}

The best-fit parameters for our MDFs are $\gamma=1.14_{-0.16}^{+0.22}$, $\delta=1.74_{-0.09}^{+0.15}$, $\mathrm{loc}=0.22_{-0.03}^{+0.04}$, $\mathrm{scale}=0.28_{-0.02}^{+0.02}$ for FGK-type stars, and $\gamma=0.47_{-0.34}^{+0.18}$, $\delta=1.68_{-0.25}^{+0.37}$, $\mathrm{loc}=0.09_{-0.05}^{+0.03}$, $\mathrm{scale}=0.20_{-0.04}^{+0.06}$ for M-type stars. We plot our fitting results in \Cref{fig:fgkm_mdfs}. 

In order to obtain the overall metallicity distribution for stars in the $M_{\mathrm{G}}$ range from 4.1 to 12.1 mag, we conduct a weighted average between the intrinsic MDFs of FGK-stars and M-stars. The weights should be their respective numbers of selection-function-resampled objects in the same distance range, which is about 1 versus 2.6 for our FGK- and M-dwarf samples in their overlapped distance of 100-160 pc. Thus, the parameters that we apply to depict the intrinsic metallicity distribution function in the solar neighbourhood are $\gamma=0.46$, $\delta=1.48$, $\mathrm{loc}=0.09$, $\mathrm{scale}=0.20$.

\section{Measurement for the Present-day Mass-ratio Distribution in the Solar-neighbourhood with Wide Binaries} \label{sec:app_q}

As stated before, we could not constrain the mass-ratio distribution with GCNS since it is a mixture of different stellar population and the information for mass-ratio distribution might be wiped out. Therefore, we fix the present-day mass-ratio distribution with some reasonable parameter value in our model. We use the wide binary catalog published by \citet{el-badryMillionBinariesGaia2021} to derive the present-day mass-ratio distribution. Selection effect should be clearly considered during this derivation. The completeness for wide binaries with narrower separations decreases more significantly when the distance goes up, so we should balance between the farthest distance for the sample and the number of sample to assure wider range of separation we cover and the significance for the signal. We make following cuts:

(1) If we adopt the \textsl{Gaia} angular resolution with 1.5 arcsec as stated in \citet{collaborationGaiaEarlyData2021}, we can guarantee that wide binaries with projected angular resolution larger than 100 AU can be resolved if we restrain the distance to be smaller than 60 pc. So the cuts will be $\varpi<=16.7~\mathrm{mas}$ and \texttt{sep\_AU}>100 AU.

(2) To reduce the occurrence of chance alignment, we make the cut with the parameter given by \citet{el-badryMillionBinariesGaia2021}, \texttt{R\_chance\_align}<=0.75, which is the value removing most chance alignments according to their investigation. Since we restrict the distance to a small volume, this relatively loose cut on \texttt{R\_chance\_align} is sufficient. This cut will remove about 1\% of our sources.

\begin{figure*}
    \centering
    \includegraphics[width=0.8\linewidth]{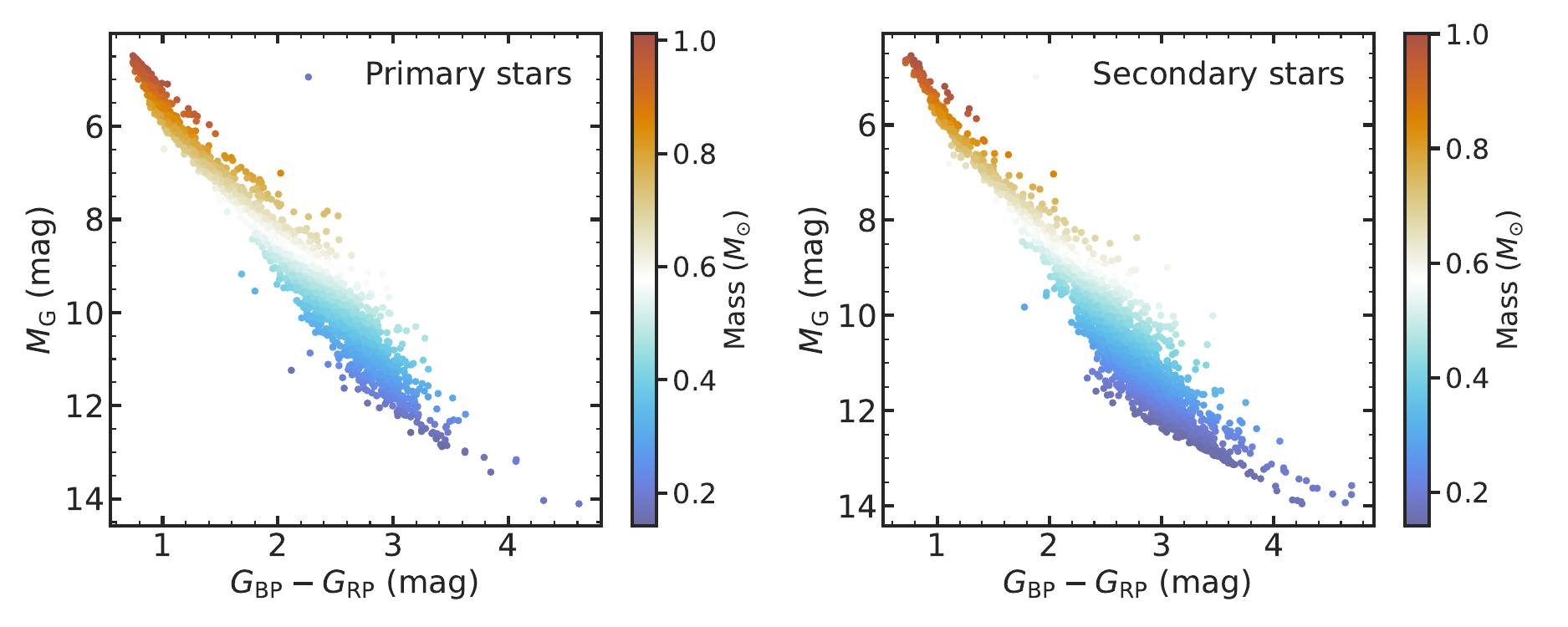}
    \caption{Masses obtained by interpolation of PARSEC isochrones of primary and secondary stars for wide binary systems with $\varpi<=16.7~\mathrm{mas}$ and $\mathrm{sep}>100~\mathrm{AU}$ from \citet{el-badryMillionBinariesGaia2021}'s catalog.}
    \label{fig:wb_mass}
\end{figure*}

\begin{figure}
    \centering
    \includegraphics[width=0.8\linewidth]{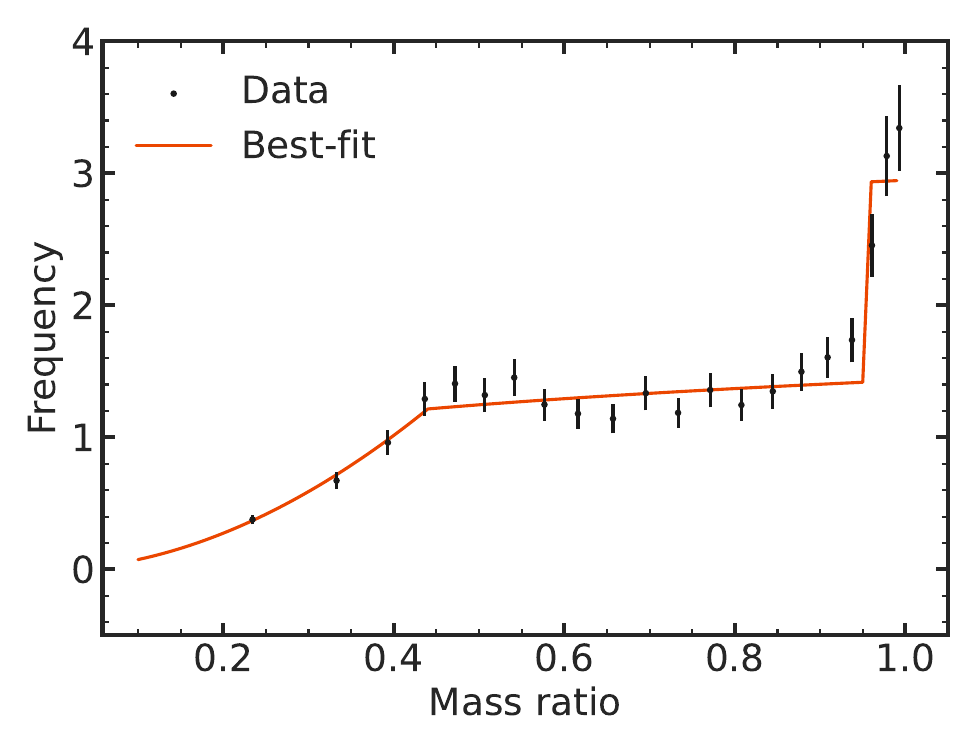}
    \caption{Mass-ratio distribution for wide binary systems with $\varpi<=16.7~\mathrm{mas}$ and $\mathrm{sep}>100~\mathrm{AU}$. The red solid line shows our best fitting result with $\gamma_1=1.89_{-0.20}^{+0.23}$, $\gamma_2=0.20_{-0.13}^{+0.12}$, $q_{\mathrm{break}}=0.44_{-0.02}^{+0.02}$,  $F_{\mathrm{twin}}=0.08_{-0.01}^{+0.01}$, and $C=0.93_{-0.02}^{+0.02}$.}
    \label{fig:wb_q}
\end{figure}

\begin{figure*}
    \centering
    \includegraphics[width=0.8\linewidth]{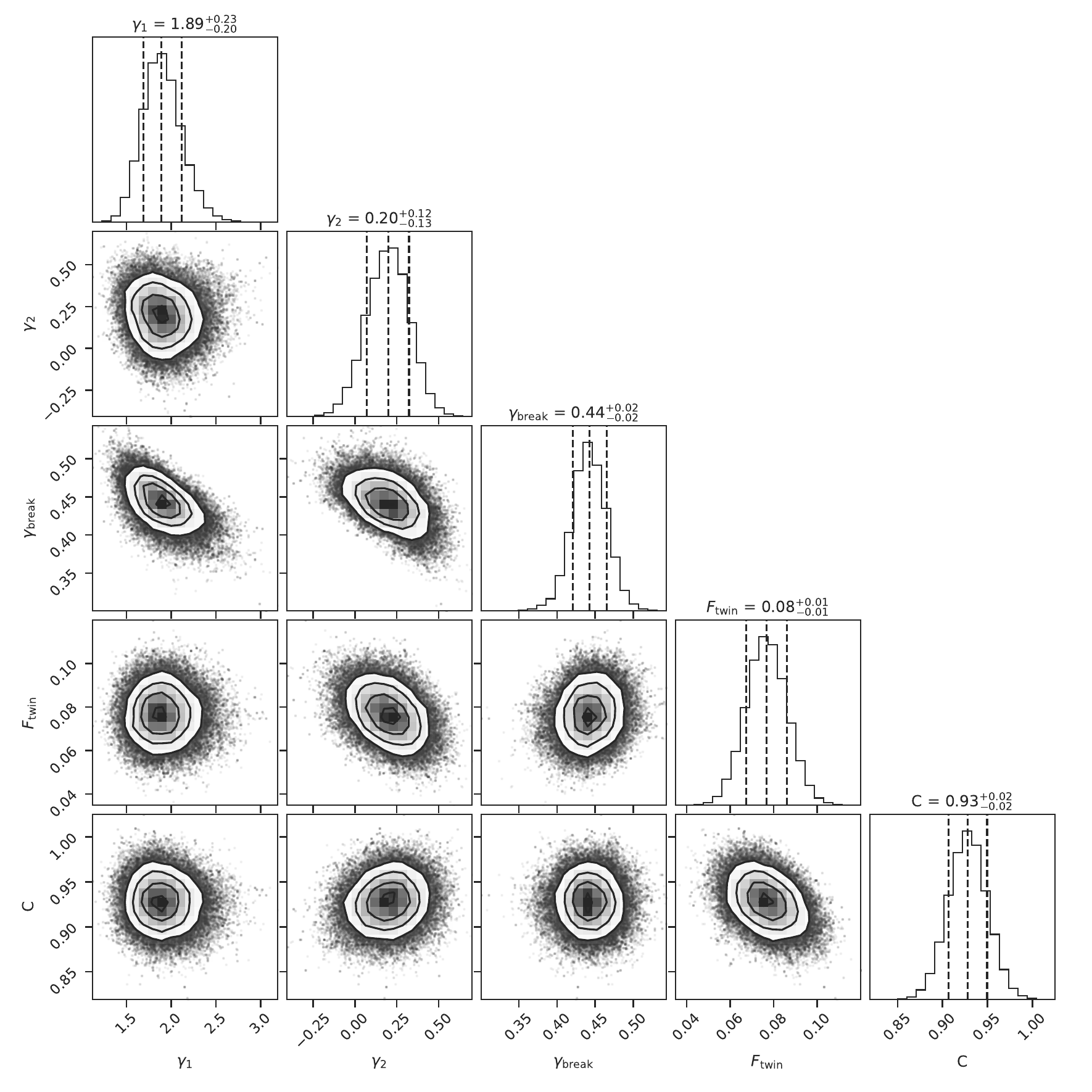}
    \caption{Posterior distribution for the parameters modeling the wide binary mass-ratio distribution.}
    \label{fig:wb_q_mcmc}
\end{figure*}

At last we obtain approximately $2,100$ pairs of wide binaries. We use the PARSEC model version 1.2s to transform brightness and colour to mass and metallicity, which is demonstrated in \Cref{fig:wb_mass}, and then we can obtain the mass-ratio distribution in \Cref{fig:wb_q}. We parametrize this distribution following \citet{moe2017}, which includes a broken power-law to describe the distribution at $0.2<q<0.95$ with power-law indices $\gamma_1$ and $\gamma_2$, and the break point $q_{\mathrm{break}}$, and includes the twin binary fraction $F_{\mathrm{twin}}$. The best-fitting result is displayed in \Cref{fig:wb_q_mcmc} with $\gamma_1=1.89_{-0.20}^{+0.23}$, $\gamma_2=0.20_{-0.13}^{+0.12}$, $q_{\mathrm{break}}=0.44_{-0.02}^{+0.02}$,  $F_{\mathrm{twin}}=0.08_{-0.01}^{+0.01}$, and $C=0.93_{-0.02}^{+0.02}$ which is a normalization parameter for the mass-ratio probability function.

We should notice that in practice we use this distribution to represent the mass-ratio distribution for the unresolved binaries in the solar neighbourhood. So here we make a strong assumption that the mass-ratio distribution for binaries with separation wider than 100 AU is representative for all binaries.

%%%%%%%%%%%%%%%%%%%%%%%%%%%%%%%%%%%%%%%%%%%%%%%%%%

% Don't change these lines
\bsp	% typesetting comment
\label{lastpage}
\end{document}